\documentclass[3p]{scrartcl}

\usepackage{titling}
\usepackage[hmarginratio=1:1,top=32mm,left=15mm,columnsep=1cm]{geometry} %
\usepackage{array}
\usepackage{color} 
\usepackage{tabularx}
\usepackage{graphicx}
\usepackage{subfig}
\usepackage[svgnames,table,xcdraw]{xcolor}
\usepackage{amsmath}
\usepackage{amssymb}
\usepackage{amsfonts}	
\usepackage{comment}
\usepackage{multirow}
\usepackage{relsize} 
\usepackage{fourier} 
\usepackage{epstopdf}
\usepackage[font=small,labelfont=bf]{caption}	
\usepackage{multicol}

\usepackage{nomencl}	
\usepackage{pbox} 
\usepackage{xcolor,cancel}

\makenomenclature

\usepackage[backend=bibtex,giveninits=true,sorting=nyt,url=false,doi=true,eprint=false,isbn=false,
backref,backrefstyle=none,maxbibnames=99]{biblatex}
\DefineBibliographyStrings{english}{%
	backrefpage = {Cited on p\adddot},%
	backrefpages = {Cited on pp\adddot}%
}
\bibliography{references}

\renewbibmacro{in:}{%
	\ifboolexpr{%
		test {\ifentrytype{article}}%
	}{}{\printtext{\bibstring{in}\intitlepunct}}%
}

\newcommand\BibTeX{{\rmfamily B\kern-.05em \textsc{i\kern-.025em b}\kern-.08em
T\kern-.1667em\lower.7ex\hbox{E}\kern-.125emX}}

\allowdisplaybreaks[1]

\usepackage{hyperref} 
\hypersetup{
    colorlinks=true,                          
    linkcolor=Blue, 
    citecolor=DarkRed, 
    urlcolor=blue  } 

\newcommand{\pd}{\partial}
\newcommand{\crho}{\varrho}
\newcommand{\ceps}{\varepsilon}
\newcommand{\epsnd}{\epsilon}

\newcommand{\bb}{\boldsymbol{b}}
\newcommand{\cbar}{\crho}
\newcommand{\mm}{\boldsymbol{m}}
\newcommand{\BB}{\boldsymbol{B}}
\newcommand{\DD}{\mathcal{D}}
\newcommand{\vv}{\boldsymbol{v}}
\newcommand{\ww}{\boldsymbol{w}}
\newcommand{\xx}{\boldsymbol{x}}
\newcommand{\yy}{\boldsymbol{q}}

\newcommand{\del}{\nabla}
\newcommand{\rmd}{\mathrm{d}}
\newcommand{\rmdt}{{\mathrm{d}t}}
\newcommand{\md}[1]{\frac{\rmd #1}{\rmdt}}

\newcommand{\bm}[1]{\boldsymbol{#1}}
\newcommand{\mylength}{\text{m}}
\newcommand{\mytime}{\text{s}}
\newcommand{\mymass}{\text{kg}}
\newcommand{\SigmaMod}{\Sigma^2}
\newcommand{\calO}{\mathcal{O}}

\newcommand{\rhoE}{\mathcal{E}}

\newcommand{\oomega}{\boldsymbol{\omega}}

\begin{document} 

\title{ \Large
	\textbf{Nonequilibrium model for compressible two-phase 
	two-pressure flows with surface tension} 
}
\author{
Ilya Peshkov
\thanks{Department of Civil, Environmental and Mechanical Engineering, 
University of Trento, Via Mesiano 77, Trento 38123, Italy
\href{mailto:ilya.peshkov@unitn.it}{ilya.peshkov@unitn.it}
}
\quad
Evgeniy Romenski\thanks{Sobolev Institue of Mathematics, 4 Acad. Koptyug avenue, Novosibirsk, 
Russia, 	
	\href{mailto:author.two@email.com}{evrom@math.nsc.ru}}
\quad
Michal Pavelka\thanks{Mathematical Institute, Faculty of Mathematics and Physics, Charles University, 
Sokolovská 83, Prague 186 75, Czech Republic, 
	\href{mailto:pavelka@karlin.mff.cuni.cz}{pavelka@karlin.mff.cuni.cz}}
}
\thanksmarkseries{arabic}

%
\maketitle 
\date{\today}

\begin{abstract}
\noindent
In continuum thermodynamics, models of two-phase mixtures typically obey the condition of 
pressure equilibrium across interfaces between the phases. We propose a new non-equilibrium model 
beyond that condition, allowing for microinertia of the interfaces, surface tension, and 
different phase pressures. The model is formulated within the framework of Symmetric Hyperbolic Thermodynamically Compatible equations, and it possesses variational and Hamiltonian structures. Finally, via formal asymptotic analysis, we show how the pressure equilibrium is restored when fast degrees of freedom relax to their equilibrium values.
\end{abstract}

\section{Introduction}

When dealing with multi-fluid flows of several immiscible fluids (gas-liquid or liquid-liquid 
mixtures), one needs to take into account the effect of surface tension, for instance in
dispersed flows, bubbly fluid, sprays droplets, or (super-)fluids undergoing phase transitions \cite{hirschler2017,vitrano2022}. 
The interface between the mixture constituents can be either well-resolved or under-resolved. In the former, surface tension defines the shape of the macroscopic interface, while in the latter, it introduces microinertia due to the 
bubbles/droplets oscillations. 
In this paper, we propose a continuum mechanics model for multi-phase flows with macroscopically 
resolved or continuously distribute interfaces that adheres to both the principles of 
thermodynamics and Hamiltonian mechanics.

In continuum mechanics, there are two alternative approaches to address the effect of surface 
tension: the sharp interface and diffuse interface approaches, e.g. see 
\cite{Popinet2018,Saurel2018}. The sharp 
interface approach treats an interface as a true hypersurface of zero thickness separating pure 
phases. This is a pure geometrical approach, and any extra physics (e.g. phase transition, mass 
transfer, etc.) must be introduced as a boundary effect on the interface. On the other hand, in  
diffuse-interface-type approaches, a new state variable is introduced as a continuous field 
representing the interface as a narrow mixing zone in which all constituents coexist. 
In contrast to the sharp interface approaches, extra physics must be added via coupling of the 
interface field with other physical fields both at the governing-equation level and through the constitutive relations.
Modeling surface 
tension with a diffuse interface approach represents the subject of this paper.

In turn, inside of the diffuse interface community, the multi-fluid system can be considered from 
the 
two different perspectives. According to the first viewpoint, the mixture is treated as a single 
continuous medium (single-fluid approach) without distinguishing the individual constituents. The 
famous examples of such 
theories are the Korteweg type models and phase 
field models, e.g. \cite{Freistuhler2017b,Bresch2019,Gouin1988}, with a single double-well potential serving as the 
equation of state of the whole 
multifluid system. On the other hand, according to the second viewpoint, the multifluid system is 
considered as a mixture with 
well defined species governed by their individual parameters (velocities, pressures, temperatures, 
etc.), and most importantly, individual equations of state. It is, therefore, more realistic 
and thus has a 
bigger potential than the 
single-fluid approach because it contains physically motivated extra degrees of freedom missing in 
the latter. Moreover, the single-fluid approach is restricted to liquid-liquid or 
gas-liquid systems, while the mixture approach can be potentially also applied to the solid-fluid 
interfaces or dispersed multi-phase flows in porous elastic media \cite{Romenski2022}, in 
particular in the setting of the unified model of continuum mechanics 
\cite{HPR2016,DPRZ2016,nonNewtonian2021,SPH_SHTC}.  

In a truly non-equilibrium mixture model, all corresponding phase parameters (pressures, velocities,
 etc.) are independent and distinct within a mixture element. Phase parameters may relax to 
 common values only in the vicinity of thermodynamic equilibrium. In such instances, various 
 reduced models can be derived, such as single-pressure models and single-velocity models, as 
 described in \cite{StewartWendroff1984,Kapila2001}. Notably, the single-pressure approximation 
 is widely employed in diffuse interface surface tension models, as exemplified in 
 \cite{Brackbill1992,Saurel2005,Schmidmayer2017,Chiocchetti2021}. This raises a fundamental 
 question: how can surface tension be incorporated into a two-fluid diffuse interface model 
 without invoking the pressure-equilibrium assumption? This paper addresses this question 
 within the framework of the Symmetric Hyperbolic Thermodynamically Compatible (SHTC) class 
 of equations.

The SHTC formulation for two-phase flows was first proposed in  \cite{Rom1998,Rom2001}, and later 
it was developed in a 
series of papers \cite{Romenski2004,Romenski2007,RomDrikToro2010}. In particular, it was generalized to
an arbitrary number of phases in \cite{Romenski2016} and it's variational and Hamiltonian formulation via Poisson brackets were discussed
in \cite{SHTC-GENERIC-CMAT}.

Other mixture formulations for multifluid systems without surface tension exist. In the community 
of compressible 
multiphase flows, perhaps the most popular is the Baer-Nunziato (BN) model \cite{BaerNunziato}. The relation 
between the BN model and the SHTC mixture model was discussed for example in 
\cite{Romenski2004,Romenski2007}. In particular, in contrast to the SHTC model, the BN model has 
the known issue of closure relations for interfacial quantities (interfacial velocity and 
pressure), which is linked to the lack of variational formulation for the BN model. Variational 
formulations for binary mixtures were also proposed in \cite{Gavrilyuk2002,GouinRuggeri2003}. 
The model proposed by Ruggeri in \cite{Ruggeri2001,GouinRuggeri2003} only applies to homogeneous 
mixtures (no volume fraction) and the interactions between phases reduced to interfacial friction 
only (no pressure relaxation, no temperature relaxation) and therefore a direct comparison between 
the SHTC formulation and \cite{Ruggeri2001} is impossible at the moment.

On the other hand, the model proposed by Gavrilyuk and Saurel in \cite{Gavrilyuk2002} was directly 
designed to address the 
micro-capillarity and microinertia effects in bubbly liquids. 
Thus, it contains the volume fraction as a state variable and describes, like the BN model, 
a two-fluid mixture as a medium with two pressures and two velocities.
The SHTC model for surface tension discussed in this paper is very close, in principle, to 
the one in \cite{Gavrilyuk2002}. By this we mean that the time and space gradients of the volume 
fraction are introduced in our model as new state variables to account for the microinertia and 
mixture heterogeneity in the vicinity of the interface. However, the different choice of the time 
gradient of the volume fraction 
results in overall different governing equations in the two approaches. It is likely that model 
\cite{Gavrilyuk2002} can also be applied not only to modeling of the microinertia effect in bubbly 
fluids, but also to describe macroscopic diffuse interfaces between immiscible fluids as in this paper, but this option has not yet been tested for 
\cite{Gavrilyuk2002}. 

Finally, we would like to emphasize the variational nature of the SHTC equations in general, and 
the SHTC formulation for mixtures in particular. When dealing with multiphysics problems it is 
important that the coupling of various physics in the system of governing equations is done in a 
compatible way. Human intuition can not serve as a reliable tool in the derivation of governing 
equations, but one should use first-principle-based approaches. The SHTC equations discussed in this 
paper, can be derived by two first-principle-type means. The first is the variational principle 
which is discussed in Sec.\,\ref{sec.app.VarPrinciple}. We also demonstrate that the governing 
equations can be derived from the Hamiltonian formulation of non-equilibrium thermodynamics known 
as GENERIC (General Equations for Non-Equilibrium Reversible-Irreversible Coupling) 
\cite{go,og,hco,PKG_Book2018}. Thus, the discussed SHTC equations for surface tension admits a variational 
and Hamiltonian formulation via non-canonical Poisson brackets.

The paper is organized as follows. First, in Sec.\,\ref{sec.SHTC.twofluid} we briefly recall the 
SHTC model for binary heterogeneous 
mixtures. Then, in Sec.\,\ref{sec.SHTC.ST}, we discuss how it can be generalized to include the 
surface tension effect so that 
the generalization still stays in the class of thermodynamically consistent systems of first-order hyperbolic equations.
The 
variational and Hamiltonian formulations of the SHTC surface tension model are discussed in 
Sec.\,~\ref{sec.app.VarPrinciple} and \ref{sec.generic}. The closure problem is discussed in 
Sec.\,\ref{sec.closure}, 
and hyperbolicity of the governing equations is partially discussed in Sec.\,\ref{sec.hyperb}. We 
then derive the relaxation limit of the governing equation in Sec.\,\ref{sec.Asympt}, and 
demonstrate that the stationary bubble solution is compatible with the Young-Laplace law in 
Sec.\,\ref{sec.bubble}. Finally, we analyze the dispersion relation of the model in 
Sec.\,\ref{sec.dispers}.

%
%

\section{SHTC master system for two-phase compressible flows}\label{sec.SHTC.twofluid}

The governing equations of the discussed later continuous surface tension model generalize the SHTC 
equations of compressible two-phase flows \cite{Romenski2007}. Like all SHTC models, the two-phase 
flow model with surface tension can be obtained from a master SHTC system of balance equations 
\cite{GodRom1996b,GodRom1998,Rom2001} using generalized internal energy as thermodynamic potential. 
In turn, the master SHTC system can be derived from a variational principle and thus consists of a 
set of Euler-Lagrange equations coupled with trivial differential constraints. It describes 
transport of abstract scalar, vector, and tensor fields. 

Let us first remind the SHTC two-phase flow model and then we demonstrate how it can be  to account for the effect of surface tension.  

\subsection{Two-phase flow master system}

The master SHTC system for the mixture of two ideal fluids can be found in \cite{Romenski2007,RomDrikToro2010} and reads as
\begin{subequations} \label{MasterSys}
\begin{align}
 \frac{\partial \rho \alpha}{\partial{t}} &+
\frac{\partial \rho \alpha  v_{k}}{\partial{x}_{k}} = -\frac{1}{\lambda} {\rho E_{\alpha}}, 
 \label{irs1}
 \\
 \frac{\partial \rho }{\partial{t}} &+
\frac{\partial \rho v_{k}}{\partial{x}_{k}}= 0,
\label{irs2}
\\
 \frac{\partial \rho v_i }{\partial{t}} &+
\frac{\partial (\rho v_i v_{k} + \rho^2 E_\rho \delta_{lk} + \rho
w_l E_{w_k})}{\partial{x}_{k}}= 0, 
\label{irs3} 
\\
 \frac{\partial \rho c  }{\partial{t}} &+ 
 \frac{\partial(\rho c v_{k}  + \rho E_{w_k})}{\partial{x}_{k}} = 0, 
\label{irs4} 
\\
 \frac{\partial  w_k }{\partial{t}} &+ 
 \frac{\partial (v_{l}w_l + E_{c })}{\partial{x}_{k}} + 
v_l\left(\frac{\partial w_k}{\partial x_l}-\frac{\partial w_l}{\partial x_k} \right)= 
-\frac{1}{\chi} {E_{w_k}},
\label{irs5}
\\
\frac{\partial \rho S}{\partial{t}} &+
\frac{\partial \rho S v_{k}}{\partial{x}_{k}} =
\frac{\rho}{E_S}\left( \frac{1}{\lambda}  E_{\alpha} E_{\alpha} + \frac{1}{\chi} E_{w_k}E_{w_k} 
\right) \geq 0.
\label{irs6} 
\end{align}
\end{subequations}
This system describes a mixture whose infinitesimal element of volume $V=V_1 + V_2$ and mass $M = 
M_1 + M_2$ is characterized by the following state 
variables. Here, $V_a$, $M_a$, $a=1,2$  are the volume and mass of the constituents of the mixture in the volume $V$. 
The mass density of the mixture is defined as $\rho = M/V$, while the apparent densities of the 
constituents in the volume $V$ are $\crho_a = M_a/V$, $a=1,2$, so that $\rho = \crho_1 + \crho_2$. 
Also, the non-dimensional 
mixture parameters are the mass $c_a = M_a/M = \crho_a/\rho$  and volume $\alpha_a = V_a/V $ 
fractions. Note that from
these definitions it follows that $c_1+c_2=1$ and $\alpha_1 + \alpha_2 = 1$ and hence one needs to know only the mass
and volume fraction of one of the components. Thus, in the above equations, we use the notations $c = c_1$ 
and $\alpha=\alpha_1$. 

Furthermore, one needs the so-called actual densities $\rho_a = M_a/V_a $ of the phases to define the individual
equations of state (internal energies) $e_a(\rho_a,s_a)$ of the constituents, where $s_a$, $a=1,2$ are the specific 
entropies of the constituents, and $\rho S = \crho_1 s_1 + \crho_2 s_a$ is the mixture entropy 
density. One can observe the relation $\crho_a = \alpha_a \rho_a$, $a=1,2$.
Despite the original SHTC two-flui model \cite{Romenski2007,RomDrikToro2010} is a two-temperature 
model, in this paper we ignore some thermal properties of such mixtures, and a single 
entropy-approximation $s_1 = s_2 = S$ is adopted for the sake of simplicity.

To describe the kinematics of the mixture element we define the mixture momentum, which is the sum 
of the constituents' momenta (due to the momentum conservation principle) $\rho \vv = 
\crho_1 \vv_1 + \crho_2 \vv_2$ with $\vv_a = {v_{a,k}}$ ,$a=1,2$, $k=1,2,3$ being the velocity 
of the $a$-th component. Note that the velocity of the mixture element is thus defined as 
\begin{equation}
	\vv = \frac{\crho_1}{\rho} \vv_1 + \frac{\crho_2}{\rho}\vv_2 = c_1 \vv_1 + c_2 \vv_2.
\end{equation}
Additionally, we introduce the relative velocity $\ww=\vv_1 - \vv_2$ which is defined with respect
to the second phase but not with respect to the mixture velocity $\vv$ as traditionally done, see 
for example \cite{Ruggeri2015}. This is required by the structure of the SHTC equations and 
eventually by 
the variational scheme we use \cite{SHTC-GENERIC-CMAT}.

One can notice that the fluxes and sources of the governing equations are defined in terms of the 
partial derivatives of the mixture total energy $E = E(\rho,S,\alpha,c,\ww,\vv)$ with respect to 
the state variables, e.g. $E_{\rho} = \frac{\partial E}{\partial 
\rho}$, $E_{\alpha} = 
\frac{\partial E}{\partial \alpha}$, $E_{w_k} = \frac{\partial E}{\partial w_k}$, the so-called 
thermodynamic forces.

\subsection{First law of thermodynamics}


Thermodynamic consistency of the SHTC equations means that the first and second laws of 
thermodynamics are satisfied by constructions. Indeed, it can be shown that on the solution to 
\eqref{MasterSys} an additional conservation law
\begin{equation}\label{EnergyConserv} 
\frac{\partial \rho E}{\partial t}  + 
\frac{\partial }{\partial x_k} \left( \rho v_k E + v_i\left( \rho^2 E_\rho \delta_{ki} + \rho w_i 
E_{w_k} \right) + 
\rho E_c E_{w_k} \right) =0,
\end{equation}
is satisfied. It can be obtained by multiplying equations of \eqref{MasterSys} by corresponding 
multipliers and summing all them up, see \cite{Rom1998,Rom2001,SHTC-GENERIC-CMAT}.
Here, $\rho E = \rho E(\rho,S,\alpha,c,\vv,\ww)$ is the total energy density of the mixture which, 
due to the energy conservation principle, 
is nothing else 
but the sum $\rho E = \crho_1 E_1 + \crho_2 E_2$ of the total energies $E_a = e_a(\rho_a,S) + 
\frac{1}{2}\vv_a^2$, $a=1,2$ of the 
constituents. After a certain term rearrangements, the specific mixture energy $E = c_1 E_1 + c_2 
E_2$ 
can be written in terms of the SHTC state variables
\begin{equation} \label{eqn.energy.2phase}
	E = c_1 e_1(\rho_1,S) + c_2 e_2(\rho_2,S) + c_1 c_2 \frac{1}{2} \ww^2 + \frac{1}{2} \vv^2.
\end{equation}

The presence of an additional conservation law like \eqref{EnergyConserv} makes it possible to 
reformulate \eqref{MasterSys} 
in terms of the so-called generating thermodynamic potential and new variables (thermodynamically 
dual to the original ones) and to transform the equations to a symmetric form, e.g. see 
\cite{SHTC-GENERIC-CMAT}. If the potential $\rho E$ is convex in terms of the state variables, then 
the system is symmetric hyperbolic and that is why the name (SHTC) of the equations. 

To close system \eqref{MasterSys}, one needs to provide the so-called closure, which for the SHTC 
equations is 
always the energy potential $\rho E$. In the case of mixtures of ideal fluids, in 
\eqref{EnergyConserv}, 
we need only to specify the internal energies $e_a(\rho_a,S)$ of the constituents. Note that the 
actual densities $\rho_a$ do not belong to the set of SHTC state variables but must be expressed 
as $\rho_a = \rho c_a/\alpha_a$. This can be used to compute all the partial derivatives $E_\rho$, 
$E_\alpha$, $E_c$, etc. in \eqref{MasterSys} 
\begin{subequations}\label{ThermForces}
	\begin{align} 
		\frac{\partial E}{\partial \rho} &= \frac{ \alpha_1 p_1+\alpha_2 p_2}{\rho^2} = \frac{ 
		\alpha p_1+(1-\alpha) p_2}{\rho^2},
		\\
		\frac{\partial E}{\partial c} &= \mu_1 - \mu_2 + (1-2c)\frac{\ww^2}{2},
		\\
		\frac{\partial E}{\partial \alpha} &= -\frac{p_1-p_2}{\rho},
		\\
		\frac{\partial E}{\partial {w_k}} &= c_1c_2 w_k = c (1-c) w_k,
	\end{align}
\end{subequations}
where $ p_a=\rho_a^2 \frac{\partial e_a}{\partial \rho_a}, a=1,2 $ are the phase pressures, and 
$\mu_a = e_a + \frac{p_a}{\rho_a} - s_a T_a = e_a - \alpha_a \frac{\partial e_a}{\partial \alpha_a} 
- s_a \frac{\partial e_a}{\partial s_a}$ is the chemical potential of the $a$-th constituent.

Finally note, that the mixture thermodynamic pressure is defined as
\begin{equation}
	p = \rho^2 E_\rho = \alpha_1 p_1 + \alpha_2 p_2.
\end{equation}

\subsection{Second law of thermodynamics and irreversibility}

For the sake of simplicity, we ignore the various dissipative processes such as heat conduction, 
viscous dissipation, phase transformation, etc., but we keep only two dissipative process in this 
consideration which are related to the main subject of the paper that is how to introduce surface 
tension in the SHTC two-phase flow model. 

The first dissipative process is the relaxation of the phase velocities to a common value, which is 
described by a relaxation-type source term $\chi^{-1}E_{w_k} = \chi^{-1} c(1-c) w_k$ in the 
relative velocity equation \eqref{irs4} with $\chi$ being a relaxation parameter with the 
dimension of time.
The second and the most important relaxation process in the context of surface tension is the 
pressure relaxation towards a common pressure which is modeled by the relaxation source term 
$\lambda^{-1}\rho E_\alpha = 
\lambda^{-1}(p_1-p_1)$ with $\lambda$ being usually a small parameter that controlls the rate of 
relaxation.

As it is seen from these two examples of dissipative processes, the dissipative terms in the SHTC 
equations (including other SHTC models for heat conduction or viscous dissipation 
\cite{HPR2016,DPRZ2016,DPRZ2017,SHTC-GENERIC-CMAT}) 
are algebraic relaxation-type terms that have the form of the so-called gradient dynamics 
\cite{SHTC-GENERIC-CMAT,PKG_Book2018}, i.e. 
they are proportional to the anti gradients (in the space of state variables) of the total energy 
(thermodynamic forces) with positive factors that control the rate of dissipation. Thus, in the 
state space, the dissipation is directed towards diminishing the thermodynamic forces $E_\alpha$, 
$E_{w_k}$, etc.

Modeling the dissipation via the gradient dynamics, we automatically guarantee the 
consistency with the first and second law of thermodynamics. Indeed, the presence of dissipative 
sources on the right-hand side does not violate the conservation of the total energy, i.e. the 
energy conservation law has a zero on the right hand-side. Of course this is achieved by the proper 
choice of the entropy production term in the entropy equation \eqref{irs6} which is canceled out 
with the rest of the dissipative source terms and simultaneously is staying always non-zero by 
construction, see more details in \cite{Romenski2007,RomDrikToro2010,SHTC-GENERIC-CMAT}.

\section{Nonequilibrium SHTC formulation of surface tension}\label{sec.SHTC.ST}

System \eqref{MasterSys} allows for resolution of macroscopic interfaces between the phases in a 
diffuse interface manner
\cite{Anderson1998,Saurel2018} using the volume fraction $\alpha$ as the 
so-called color function. Yet, such interfaces are only passively advected by the flow and do not 
carry any energy content (zero surface energy). In other words, the surface tension effects can not 
be modeled with the two-phase flow model \eqref{MasterSys}. 

In the diffuse interface setting, there several approaches exist that allow inclusion of the 
surface tension. All of them require computation of the gradient $\del \alpha = \left\{\frac{\pd 
\alpha}{\pd x_k}\right\}$ of the volume 
fraction or another smooth scalar field $\phi$ generally called a \textit{color function}. Roughly 
speaking, 
we can divide these diffuse interface approaches to surface tension into two categories: 
equilibrium models and nonequilibrium ones. In an equilibrium model, the continuous equivalent of 
the Young-Laplace law
\begin{equation}\label{eqn.YL}
	\left[p\right] = \sigma \kappa, \qquad \kappa:= \nabla\cdot \left( \frac{\nabla\phi}{\Vert 
	\nabla\phi\ \Vert 
	}\right) 
\end{equation}
is directly used as the constitutive relation for the stress tensor. In \eqref{eqn.YL}, $\sigma$ is 
the surface tension coefficient, and square brackets $\left[ \bullet \right] $ denote the jump of a 
quantity across the interface, in particular the jump of the pressure $p$ in \eqref{eqn.YL}. This 
law is 
known 
to be a good approximation for interfaces not far 
from mechanical and/or thermodynamic equilibrium. Representatives of the equilibrium approach are 
the models that for example can be found in 
\cite{Brackbill1992,Saurel2005,Schmidmayer2017,Chiocchetti2021}.

On the other hand, in a non-equilibrium surface tension model, the Young-Laplace law is not 
directly prescribed as a constitutive function but is recovered if the flow is not far from the 
mechanical and thermodynamic equilibrium (typically, in the limit when a small parameter 
(capillarity coefficient) goes to 0). Representatives of the nonequilibrium approach are the phase 
field models, for 
example, the
\textit{Korteweg}-type\footnote{First idea was proposed by van der Waals in \cite{vdW1979}.} models 
\cite{Gouin1988,Dhaouadi2022,Rohde2023}, to which \textit{Cahn-Hilliard}-type models are 
closely related \cite{Cahn1958,Lowengrub2005,Freistuhler2017b}. However, as was mentioned in the 
introduction, such models belong to the so-called single-fluid-type models and thus has intrinsic 
limitations for modeling mixtures far from thermodynamic 
equilibrium when the mixture constituents are having different state parameters, e.g. pressures, 
temperatures, velocities, etc. 

The SHTC formulation of the surface tension we shall discuss in what follows belongs to the 
non-equilibrium type models, yet it of course has some conceptual differences from the 
Korteweg-type 
formulations as being a multi-fluid-type formulation. We also remark, that the presence of 
gradients of the state variables in the 
constitutive relations, like in \eqref{eqn.YL}, is not allowed in the SHTC theory which includes 
only first-order 
hyperbolic equations. Therefore, any space or time gradients of the fields must be lifted to the 
role of new independent state variables with their own time evolution equations.

The key state variable in representing the interfaces between the phases is the volume 
fraction $\alpha$. In a mixture element, the volume fraction can change both due to changes in 
the thermodynamic state of the constituents (e.g. thermal expansion) and when a phase flows in and 
out of the mixture element. The later is taken into account in \eqref{irs1} via simply the 
advection terms on the left-hand side, while the former is taken into 
account via the dissipative pressures relaxation source term. For the further discussion it is 
convenient to rewrite the balance law \eqref{irs1} in the following equivalent form 
\begin{equation}\label{VFeqn}
	\frac{\partial \alpha}{\partial{t}} +
	v_j \frac{\partial \alpha}{\pd x_j} = -\frac{1}{\lambda} { E_{\alpha}}.
\end{equation}

Obviously, at the mechanical equilibrium ($\pd/\pd t = 0$ and $\vv=0$), or sufficiently close to 
the thermodynamic equilibrium, i.e. if the time scale $t_\lambda$ associated with the relaxation 
parameter $\lambda$ is significantly smaller $t_\lambda/t_0 \ll 1$ than the flow time scale $t_0$, 
the 
pressure 
relaxation term $\lambda^{-1}E_{\alpha} = (p_2-p_1)/(\rho\lambda)$ drives the mixture element to a 
state with the vanishingly small pressure difference $\left[p\right] = p_1 - p_2 \approx 0 $. This 
fact is, of course, not compatible with the 
Young-Laplace law \eqref{eqn.YL}. Therefore, the volume fraction evolution equation must be 
subjected to some modifications.

Let us consider the balance equation for the volume fraction in a more general than \eqref{VFeqn} form
\begin{equation}\label{VFeqnG}
\frac{\partial \alpha}{\pd t} +
 v_j \frac{\partial \alpha}{\pd x_j} =- \mathcal{F},
 \end{equation}
where the algebraic source term $\mathcal{F}$ has to be defined.
From \eqref{VFeqnG} one immediately can obtain an equation for the new vector field $\bb 
:= \del\alpha$ by differentiating \eqref{VFeqnG} with respect to $x_k$:
\begin{equation} \label{b_eqn}
\frac{\partial b_k}{\pd t} +
 v_j \frac{\partial b_k}{\pd x_j} + b_j \frac{\partial v_j}{\pd x_k} +
 \frac{\partial \mathcal{F}}{\pd x_k}  =0,
\end{equation}
or in an equivalent form
\begin{equation} \label{b_eqn1}
\frac{\partial b_k}{\pd t} + \frac{\pd \left(b_j v_j + \mathcal{F}\right)}{\pd x_k} +
 v_j \left( \frac{\partial b_k}{\pd x_j}-\frac{\partial b_j}{\pd x_k} \right)  = 0,
\end{equation}
where the source term $\mathcal{F}$ from \eqref{VFeqnG} has become the constitutive flux in 
\eqref{b_eqn1}.

Note that, due to its definition, $\bb$ must satisfy the stationary differential curl-constraint
\begin{equation} \label{curl}
	\frac{\partial b_k}{\pd x_i}-\frac{\partial b_i}{\pd x_k} =0,
\end{equation}
that if holds for the initial data must remain so for all later times. This also should hold at the 
discrete level when solving \eqref{b_eqn1} numerically. One may expect, however, that condition 
\eqref{curl} 
could be dropped in the supercritical case, when the difference between liquid and gas phases 
disappears.

Let us remark that in the traditional approach to continuous surface tension modeling 
\cite{Brackbill1992}, see also\cite{Saurel2005,Schmidmayer2017,Chiocchetti2021}, the hypothesis of 
equal phase pressures $p_1 = p_2$, or $E_\alpha = 0$, (single pressure approximation) is employed 
from the very
beginning. In other words, the curvature $\kappa = \del\cdot\left( \frac{\del\phi}{\Vert \del \phi 
\Vert} \right)$ 
of the interface 
is computed from the color function $\phi$ governed by the pure transport equation
\begin{equation}\label{eqn.color.fun}
	\frac{\pd \phi}{\pd t} + v_j \frac{\pd \phi}{\pd x_k} = 0
\end{equation}
which is equivalent to the homogeneous equation 
\eqref{VFeqnG} with $\mathcal{F} = 0$. However, far from thermodynamic equilibrium, the phase 
pressures are different $p_1\neq p_2$, and so are the gradients of the color function $\del\phi$  
and of the volume 
fraction $\del\alpha$ computed from \eqref{eqn.color.fun} and \eqref{VFeqnG}, accordingly.

Thus, equation \eqref{b_eqn1} is the starting point for formulating the SHTC governing equations 
for 
surface tension.
In accordance with the SHTC formalism, the governing equations have a pair structure, or a 
\textit{Hamiltonian structure} \cite{GodRom1998,SHTC-GENERIC-CMAT,PPK2020}, that is the governing 
equations are split into pairs and in each pair, one equation is an Euler-Lagrange equation 
and the second equation is a differential identity (differential constraint), see 
Section\,\ref{sec.app.VarPrinciple}. Equation 
\eqref{b_eqn1} is apparently a differential identity for $\del\alpha$, and thus it must have a 
complementary Euler-Lagrange equation for a new scalar field, say $d$. This can be demonstrated 
more rigorously by the variational scheme employed in the SHTC formalism given in 
Appendix\,\ref{sec.app.VarPrinciple}, see also details in \cite{SHTC-GENERIC-CMAT,PPK2020}. 

Thus, rewriting equations \eqref{eqn.SHTC.Lagr2} from Appendix\,\ref{sec.app.VarPrinciple} in the 
Eulerian frame of reference, we obtain the following system of governing equations on the unknowns 
$\{\alpha,d,b_k\}$
\begin{subequations}\label{eqn.PDE.adb}
	\begin{align}\label{VFeqn1}
		\frac{\pd \rho\alpha}{\pd t} &+
		\frac{\pd \rho\alpha v_{k}}{\pd x_{k}} = - \rho E_d,
		\\[2mm]
		\frac{\pd \rho d}{\pd t} &+ \frac{\pd \left(\rho d v_k+\rho E_{b_k} \right)}{\pd x_{k}} 
		=\rho E_\alpha - \lambda \rho E_d,\label{d_eqn}\\[2mm]
		\frac{\pd b_k}{\pd t} &+ 
		\frac{\pd \left(b_j v_j+E_d \right)}{\pd x_k}+
		v_{j} \left( \frac{\pd b_k}{\pd x_j}-\frac{\pd b_j}{\pd x_k} \right)  =0,\label{b_eqn2}
\end{align}
\end{subequations}
where the relaxation parameter $\lambda$ is exactly the same as in \eqref{VFeqn}. Moreover, the 
source term $ \lambda \rho E_d$ is of dissipative nature \cite{PPK2020}, and is thus missing 
in the variational formulation in \eqref{eqn.SHTC.Lagr2.d} but it is added afterwards in full 
consistency with the second law of thermodynamics (it contributes to the entropy production) 
\cite{SHTC-GENERIC-CMAT}. We shall comment further on this after the full SHTC equations will be 
presented, see \eqref{MasterSysSF}.

In particular, by comparing \eqref{b_eqn1} and \eqref{b_eqn2}, or \eqref{VFeqnG} and 
\eqref{VFeqn1}, we identify $\mathcal{F}=E_d = \pd 
E/\pd d$ as the 
thermodynamic force associated with the new scalar field $d$. As it is clear from its definition, 
see 
\eqref{eqn.phi.beta} and \eqref{eqn.d.def}, the field $d$ is closely associated with the time 
derivative $\frac{\pd\alpha}{\pd t} + v_k\frac{\pd\alpha}{\pd x_k}$ of the volume fraction 
$\alpha$, and hence 
it 
carries information about the microinertia of the interface field $\alpha$. 

For practical use of equations \eqref{eqn.PDE.adb}, it is convenient to use the following re-scaled 
state variables
\begin{subequations}\label{B&D}
	\begin{alignat}{2} 
	    B_k     := &\delta b_k,			\qquad && D     := \delta^{-1} d,\\[2mm]
		E_{B_k}  = &\delta^{-1}E_{b_k}, \qquad && E_{D}  = \delta E_{d},
	\end{alignat}
\end{subequations}
with $\delta$ being a scaling parameter with the units of length
\begin{equation}\label{eqn.delta}
    \delta \sim \text{Length},
\end{equation}
which, for example, can be associated with the width of the diffuse interface.

The resulting SHTC system for two-phase compressible flows with surface tension combines 
\eqref{MasterSys} and \eqref{eqn.PDE.adb} and reads
\begin{subequations}\label{MasterSysSF}
	\begin{align} 
	 &\frac{\pd \rho }{\pd t} +
	\frac{\pd \rho v_{k}}{\pd x_{k}}= 0,
	\label{master.rho}
	\\[2mm]
	 &\frac{\pd \rho v_i }{\pd t}+
	\frac{\pd \left(\rho v_i v_k + \rho^2 E_\rho \delta_{ki} + 
		\rho w_i E_{w_k} + 
		\rho B_i E_{B_k} \right)}{\pd x_{k}}= 0, 
	\label{master.mom} 
	\\[2mm]
	& \frac{\pd \rho c }{\pd t} + \frac{\pd(\rho
	c v_{k} + \rho E_{w_k})}{\pd x_{k}} = 0, 
	\label{master.crho}
	\\[2mm]
	& \frac{\pd  w_k }{\pd t}+ \frac{\pd (v_{l} w_l + E_{c})}{\pd x_{k}} + 
	v_l\left(\frac{\pd w_k}{\pd x_l}-\frac{\pd w_l}{\pd x_k} \right)= 
	-\frac{1}{\chi} {E_{w_k}},
	\label{master.w}
	\\[2mm]
	&\frac{\pd \rho\alpha }{\pd t} +
	 \frac{\pd \rho\alpha v_{k}}{\pd x_{k}} = - \frac{1}{\delta}\rho E_D, 
	 \label{eqn.final.a} \\[2mm]
	& \frac{\pd \rho D}{\pd t} + \frac{\pd (\rho D v_k+\rho E_{B_k})}{\pd x_{k}} 
	=\frac{1}{\delta} \rho E_\alpha - \frac{1}{\varepsilon} \rho E_D ,
	   \label{eqn.final.d}
	\\[2mm]
	&
	\frac{\pd B_k}{\pd t} + \frac{\pd (B_l v_l+E_D)}{\pd x_{k}}+
	 v_{l} \left( \frac{\pd B_k}{\pd x_{l}}-\frac{\pd B_l}{\pd x_{k}} \right)  =0,
	 \label{eqn.final.b}
	\\[2mm]
	&\frac{\pd \rho S}{\pd t} +
	\frac{\pd \rho S v_{k}}{\pd x_{k}} =
	\frac{\rho}{E_S}\left(\frac{1}{\chi} E_{w_k}E_{w_k} + \frac{1}{\ceps} E_{D} E_{D} \right) \geq 
	0,
	\label{rs0}
	\end{align}
\end{subequations}
where, for convenience, we introduced a new relaxation parameter
\begin{equation}\label{eqn.eps.def}
	\ceps := \lambda^{-1} \delta^2\sim\text{Time}.
\end{equation}
Also, one can see that the interface vector field $B_k$ contributes to the stress tensor via the 
term 
$\rho B_i E_{B_k}$ which, however, was not added by hands, but it emerges automatically in the 
variational and Hamiltonian formulations as shown in \cite{SHTC-GENERIC-CMAT}.

We note that the solutions to \eqref{MasterSysSF} also satisfy an extra conservation law 
\begin{equation}\label{eqn.Econs}
	\frac{\pd \rho E}{\pd t} + \frac{\pd }{\pd x_k} \left( \rho E v_k + v_i \left( \rho^2 E_\rho 
	\delta_{ki} + \rho w_i E_{w_k } + \rho B_i E_{B_k}\right) + \rho E_c E_{w_k} + \rho E_D E_{B_k} 
	\right) = 0,
\end{equation}
that is the energy conservation law. It can be shown, e.g. see 
\cite{GodRom1996b,Rom1998,Rom2001,SHTC-GENERIC-CMAT}, 
that \eqref{eqn.Econs} can be obtained by summing up all the equations with the corresponding 
coefficients:
\begin{equation}\label{eqn.sum}
	\eqref{eqn.Econs} = \rhoE_{\rho}\cdot\eqref{master.rho} + 
						\rhoE_{\rho v_i}\cdot\eqref{master.mom} + 
						\rhoE_{\rho c}\cdot\eqref{master.crho} +
						\rhoE_{w_k}\cdot\eqref{master.w} +
						\rhoE_{\rho\alpha}\cdot\eqref{eqn.final.a} +   
						\rhoE_{\rho D}\cdot\eqref{eqn.final.d} + 
						\rhoE_{B_k}\cdot\eqref{eqn.final.b} +
						\rhoE_{\rho S}\cdot\eqref{rs0},
\end{equation}
where $\rhoE = \rho E$.

System \eqref{MasterSysSF} has two types of algebraic source terms. The relaxation source terms  
$\ceps^{-1}\rho E_D$ and $\chi^{-1}E_{w_k}$ are of dissipative nature because they rise the 
entropy, which is reflected in the entropy production term. 
On the other hand, the source terms $\delta^{-1}\rho E_D$ in \eqref{eqn.final.a} and $\delta^{-1}\rho 
E_\alpha$ in \eqref{eqn.final.d} do not contribute to the entropy production and thus, are of 
non-dissipative (or reversible) nature. 
They have the opposite signs in front of them and simply cancel each other out in the summation 
\eqref{eqn.sum}. A similar ``antisymmetric'' structure of the reversible source terms can be 
observed in other SHTC models \cite{Romenski2011,PRD_Torsion2018}. We shall refer to this type of 
source terms as \textit{dispersive} because they are responsible for non-trivial dispersive properties 
of the equations as discussed in Sec.\,\ref{sec.dispers}, see also \cite{Romenski2011}.

Note that in the compressible multi-phase flow systems, one of the important processes that 
results in the change of volume fraction is the 
phase pressure relaxation \cite{Saurel2018}. In the original two-phase SHTC 
model, it is encoded in
\eqref{MasterSys} in the relaxation source term $-\lambda^{-1}\rho E_\alpha =- 
\lambda^{-1} \left(p_2 - p_1\right) $. However, it 
may look 
like we lost this important feature in the modified volume fraction equation
\eqref{eqn.final.a}. In fact, under a proper choice of relaxation parameters, the pressure 
relaxation process still drives the evolution of volume fraction implicitly. Indeed, using  
definition \eqref{eqn.eps.def}, the evolution equation for the new field $ \rho D $ can be 
rewritten as
\begin{equation}\label{eqn.D}
	\frac{\pd \rho D}{\pd t} + \frac{\pd (\rho D v_k + \rho E_{B_k})}{\pd x_{k}} 
	= \frac{\lambda}{\delta^2} \left( \frac{\delta}{\lambda}\rho E_\alpha - \rho E_D 
	\right).
\end{equation}
From this form of the source term, it is clear that if the relaxation parameters $\lambda$ and 
$\delta$ are chosen consistently such that $\ceps = \lambda^{-1}\delta^2 $ is a small parameter, 
then during the 
time evolution
\begin{equation}
	\frac{1}{\delta} \rho E_D \approx \frac{1}{\lambda}\rho E_\alpha
\end{equation}
and \eqref{eqn.final.a} tends to \eqref{irs1}. Also, see a more detailed asymptotic analysis in 
Sec.\,\ref{sec.Asympt}.

%

\section{Closure: equation of state}
\label{sec.closure}
As it is clear from system \eqref{MasterSysSF}, the fluxes and sources are defined in terms of 
the thermodynamic forces $E_\alpha$, $E_D$, $E_c$, etc., therefore, in order to close the 
system of equations, it is necessary to define the dependence of the total energy $E$ on the state variables  
$\left\{ \alpha,\rho,c,w_k,D,B_k,S \right\}$. 

Below, we discuss a simple option for $E$ that reads 
\begin{equation}\label{eqn.E1}
    E = c_1 e_1(\rho_1,S) + c_2 e_2(\rho_2,S) + 
    c_1c_2 \frac12 w_k w_k + \frac{\beta^2}{2} D^2 + \frac{\SigmaMod}{2\rho^2} B_k B_k
\end{equation}
where two additional terms, in comparison with \eqref{eqn.energy.2phase}, are added. The quadratic 
term 
$\frac{\SigmaMod}{2\rho^2} B_k B_k$ represents the surface energy with $\Sigma$ being the 
so-called 
\textit{capillarity modulus}, and $\beta$ being a modulus characterizing the microinertial 
effects. In general, $\Sigma$ could be a second-order tensor, but here we stay constrained to the isotropic case.

The explicit expression of the stress tensor in \eqref{MasterSysSF}
\begin{equation}\label{eqn.T.SHTC}
	T_{ik} =  \rho^2 E_\rho \delta_{ki} + 
	\rho w_i E_{w_k} + 
	\rho B_i E_{B_k}
\end{equation}
becomes
\begin{equation}\label{eqn.stress.E1}
	T_{ik} = \left(\alpha_1 p_1 + \alpha_2 p_2 
	 \right)\delta_{ik} 
	- 
	\frac{\SigmaMod}{\rho} \left( B_j B_j \delta_{ik} - B_i B_k \right) 
	+
	\rho c_1 c_2 w_i w_k .
\end{equation}
After comparing its surface tension part with the conventional capillary stress tensor 
\cite{Brackbill1992,Saurel2005,Schmidmayer2017,Chiocchetti2021}
\begin{equation}\label{eqn.T.classic}
	\boldsymbol{T} = p \bm{I} - \sigma \left( \Vert \del\alpha \Vert \bm{I} - 
	\frac{\del\alpha\otimes\del\alpha}{\Vert \del \alpha \Vert}\right) ,
\end{equation}
and recalling definition \eqref{B&D}, one may conclude that to recover \eqref{eqn.T.classic} from  
\eqref{eqn.stress.E1}, one needs to define $\Sigma$ from
\begin{equation}\label{eqn.Sigma.scaling1}
	\sigma = \frac{1}{\rho} \SigmaMod \delta,  
	\qquad
	\delta = \Vert \del\alpha \Vert^{-1}.
\end{equation}
Here, $\sigma$ is the surface tension coefficient from \eqref{eqn.YL}.


The thermodynamic forces $E_\alpha$, $E_c$, $E_{w_k}$, $E_D$, $E_\rho$ and $E_{B_k}$ corresponding 
to the energy
\eqref{eqn.E1} can be
explicitly expressed as
\begin{subequations} \label{ThermForcesST12}
	\begin{align}
		\frac{\pd E}{\pd \rho} 	&= \frac{\alpha_1 p_1+\alpha_2 p_2}{\rho^2} 
		- \frac{\SigmaMod}{\rho^3} B_k B_k, \\
		\frac{\pd E}{\pd \alpha}&=\frac{p_2-p_1}{\rho}, \\
		\frac{\pd E}{\pd c} 	&= \mu_1 - \mu_2 + (1-2c)\frac{w_k w_k}{2}, \\
		\frac{\pd E}{\pd {w_k}} &= c (1 - c) w_k, \\
		\frac{\pd E}{\pd D} 	&= \beta^2 D,\\
		\frac{\pd E}{\pd B_k}  	&= \frac{\SigmaMod}{\rho^2} B_k,
	\end{align}
\end{subequations}
where the phase pressures $p_a$ and chemical potentials $\mu_a$ are defined in the same way as in 
\eqref{ThermForces}.

We note that the specification of the energy potential in the form \eqref{eqn.E1} together with the 
scaling \eqref{B&D} fixes the physical units of the new quantities as
\begin{equation}
	[D]\sim \frac{\mylength}{\mytime},
	\qquad
	[\beta]\sim [-],
	\qquad
	[B_k] \sim [-],
	\qquad
	[\Sigma] \sim \frac{\mymass}{\mylength^2 \mytime},
	\qquad
	[\delta] \sim \mylength,
	\qquad
	[\ceps] \sim \mytime.
\end{equation}

We also note that in the SHTC class of equations 
\cite{GodRom1996b,Rom1998,Rom2001,SHTC-GENERIC-CMAT}, the total energy can be arbitrary physically 
motivated potential that however should additionally provide hyperbolicity or symmetric 
hyperbolicity (convex potential) of the governing equations to have a well-posed initial value 
problem for the model. 

For example, alternatively to the quadratic surface energy \eqref{eqn.E1}, one could 
consider a different surface energy in the form 
\begin{equation}\label{eqn.sqrt.B}
	\frac{\SigmaMod}{\rho}\sqrt{B_kB_k}
\end{equation}
as in the single-pressure surface tension models
\cite{Saurel2005,Schmidmayer2017,Chiocchetti2021}. 
However, in the presented non-equilibrium framework, this form of surface energy cannot be used 
because, from the non-equilibrium thermodynamic standpoint, the thermodynamic forces, e.g. 
$E_{B_k}$, must be non-constant that can be guarantied by energy potential at least quadratic in 
the corresponding state variable $B_k$. In particular, 
our consideration of the surface energy in the form \eqref{eqn.sqrt.B} showed that the corresponding thermodynamic force $\rho E_{B_k} = \SigmaMod B_k/\sqrt{B_l B_l}$ is constant for spherically symmetric interfaces. 
Hence, the use of \eqref{eqn.sqrt.B} would result in the vanishing space gradient in the radial 
direction and subsequently in $E_{\alpha} = 0$, see 
\eqref{StatD}. The latter means that the phase pressures are equal, $p_1=p_2$ that contradicts the 
intentions of our paper.

\section{Hyperbolicity and Eigenstructure}\label{sec.hyperb}

The hyperbolicity analysis of the three-dimensional equations \eqref{MasterSysSF} is a non-trivial 
task. 
Unfortunately, the analytical expressions of the eigenvalues and eigenvectors are not available in 
the general case. Note that because of the rotational invariance of the SHTC equations 
\cite{GodRom1996b}, the eigenstructure analysis can be done in the direction $x=x_1$.

Another unfortunate finding of our research is that the energy potential \eqref{eqn.energy.2phase} 
for two-phase mixture is not convex in the SHTC state variables (the conserved variables in 
\eqref{MasterSys}), at least for a two-phase mixture with ideal gas and stiffened-gas equations of 
state, e.g. see \eqref{eqn.eos}. The energy potential \eqref{eqn.E1} for the two-phase mixture with 
surface 
tension inherits the lack of convexity from potential \eqref{eqn.energy.2phase}. Therefore, despite 
being symmetrizable, system \eqref{MasterSysSF} can not benefit from the full structure of SHTC 
class of equations, i.e. it is not symmetric hyperbolic. In particular, the root of the problem of 
non-convexity is in the internal energy part $\crho_1 e_1(\crho_1,\phi_1) + \crho_2 
e_2(\crho_2,\phi_2)$ of the total energy 
potential \eqref{eqn.energy.2phase}, where $\crho_a = \rho c_a$ and $\phi_a = \rho \alpha_a$, 
$a=1,2$. Thus, it can be shown that for a mixture of two ideal gas, or 
two stiffened-gas (or their combinations) equations of state, the determinant of the Hessian of the 
internal energy is proportional to the phase pressure difference $p_1 - p_2$, and therefore, it is 
singular in the pressure equilibrium and may have negative eigenvalues which indicates that the 
internal energy is not convex. It is likely that this is also true for other equations of state.
Despite the loss of symmetric hyperbolicity (at least in the standard SHTC scheme 
\cite{SHTC-GENERIC-CMAT}), one could still investigate whether system \eqref{MasterSysSF} is just 
hyperbolic, which is a weaker condition.

Not to replace a rigorous proof of hyperbolicity but only to give some preliminary evidences in favor of that 
system \eqref{MasterSysSF} is likely hyperbolic, at least in some intervals of state variables and 
material parameters, we report here about a numerical study of hyperbolicity of \eqref{MasterSysSF}.

This analysis suggests that, similar to the two-phase SHTC system 
\eqref{MasterSys}, whose eigenstructure was 
studied in particular in \cite{Rio-Martin2023}, equations \eqref{MasterSysSF} are only weakly 
hyperbolic (two 
eigenvectors are missing) in the form as they are presented in \eqref{MasterSysSF}. However, for 
smooth solutions, system \eqref{MasterSysSF} is equivalent to its symmetrizable form, i.e. when the 
curls 
\begin{equation}\label{eqn.curls}
	\rho E_{w_j} \left( \frac{\pd w_j}{\pd x_i} - \frac{\pd w_i}{\pd x_j}\right) =0,
	\qquad
	\rho E_{B_j} \left( \frac{\pd B_j}{\pd x_i} - \frac{\pd B_i}{\pd x_j}\right) =0,
\end{equation}
of the relative velocity $\ww$ and the vector field $\BB$ are added to the mixture momentum 
equation, 
e.g. see \cite{Rom1998,SHTC-GENERIC-CMAT}, and this later form usually possesses a full basis of 
eigenvectors. This operation does not alter the eigenvalues but only allow to recover the missing 
eigenvectors.

 Note that for $\chi = c(1-c) \chi_0$, $\chi_0=const>0$, 
$\del\times \ww =0 $ for all positive times $t>0$ if it was so at $t=0$, see 
\cite{GodRom1996b,Rom1998,SHTC-GENERIC-CMAT}, while $\del\times\BB=0$ for $\delta=const$ by its 
definition. Despite $\delta$ is defined not as a constant in \eqref{eqn.Sigma.scaling1}, we shall 
show that in practical computations $\delta$ can be chosen \textit{a priori} as a constant. 
Therefore, we 
are interested in that case of $\delta=const$, and hence we assume that $\del \BB = 0$.
Note that adding \eqref{eqn.curls} to the momentum equation, does not change the eigenvalues but 
only allows to recover the missing eigenvectors.

We rewrite the homogeneous system \eqref{MasterSysSF} in an equivalent (on smooth solutions) form 
by 
adding \eqref{eqn.curls}
to the mixture momentum equation, and then rewriting the resulting system in  
a quasilinear form in the $x=x_1$-direction
\begin{equation}\label{eqn.quasilin}
	\bm{P}_t + \mathtt{A}(\bm{P}) \bm{P}_x = 0,
\end{equation}
where $\bm{P}$ is the vector of primitive variables with the components
\begin{equation}
	\bm{P} = \{v_{1,1},v_{1,2},v_{1,3},v_{2,1},v_{2,2},v_{2,3},B_1,B_2,B_3,D,\alpha,\rho_1,\rho_2\},
\end{equation}
while the matrix $\mathtt{A}(\bm{P})$ reads
\[
\tiny
\mathtt{A} = 
\left(
\begin{array}{ccccccccccccc}
 v_{1,1} & c_2 w_2 & c_2 w_3 & 0 & 0 & 0 & 0 & -\frac{2 B_2 \Sigma ^2}{\rho ^2} & -\frac{2 B_3 
 \Sigma ^2}{\rho ^2} & 0 & \frac{\left(B_2^2+B_3^2\right) \rho_{1,2} \Sigma 
 ^2}{\rho ^3} + \frac{p_{1,2}}{\rho} & \frac{\left(B_2^2+B_3^2\right) c_1 \Sigma ^2}{\rho ^2 
 \rho _1}+\frac{C_1^2}{\rho _1} & \frac{\left(B_2^2+B_3^2\right) c_2 \Sigma ^2}{\rho ^2 \rho _2} 
 \\[2mm]
 0 & v_1 & 0 & 0 & 0 & 0 & \frac{B_2 \Sigma ^2}{\rho ^2} & 0 & 0 & 0 &-\frac{B_1 B_2 \rho_{1,2} 
 \Sigma ^2}{\rho ^3} & -\frac{B_1 B_2 c_1 \Sigma ^2}{\rho ^2 \rho _1} & 
 -\frac{B_1 B_2 c_2 \Sigma ^2}{\rho ^2 \rho _2} \\[2mm]
 0 & 0 & v_1 & 0 & 0 & 0 & \frac{B_3 \Sigma ^2}{\rho ^2} & 0 & 0 & 0 & -\frac{B_1 B_3 \rho_{1,2} 
 \Sigma ^2}{\rho ^3} & -\frac{B_1 B_3 c_1 \Sigma ^2}{\rho ^2 \rho _1} & 
 -\frac{B_1 B_3 c_2 \Sigma ^2}{\rho ^2 \rho _2} \\[2mm]
 0 & 0 & 0 & v_{2,1} & -c_1 w_2 & -c_1 w_3 & 0 & -\frac{2 B_2 \Sigma ^2}{\rho ^2} & -\frac{2 B_3 
 \Sigma ^2}{\rho ^2} & 0 & \frac{\left(B_2^2+B_3^2\right) \rho_{1,2} \Sigma 
 ^2}{\rho ^3} +\frac{p_{1,2}}{\rho}& \frac{\left(B_2^2+B_3^2\right) c_1 \Sigma ^2}{\rho ^2 
 \rho _1} & \frac{\left(B_2^2+B_3^2\right) c_2 \Sigma ^2}{\rho ^2 \rho _2}+\frac{C_2^2}{\rho _2} 
 \\[2mm]
 0 & 0 & 0 & 0 & v_1 & 0 & \frac{B_2 \Sigma ^2}{\rho ^2} & 0 & 0 & 0 & \frac{B_1 B_2 \left(\rho 
 _2-\rho _1\right) \Sigma ^2}{\rho ^3} & -\frac{B_1 B_2 c_1 \Sigma ^2}{\rho ^2 \rho _1} & 
 -\frac{B_1 B_2 c_2 \Sigma ^2}{\rho ^2 \rho _2} \\[2mm]
 0 & 0 & 0 & 0 & 0 & v_1 & \frac{B_3 \Sigma ^2}{\rho ^2} & 0 & 0 & 0 & -\frac{B_1 B_3 \rho_{1,2} 
 \Sigma ^2}{\rho ^3} & -\frac{B_1 B_3 c_1 \Sigma ^2}{\rho ^2 \rho _1} & 
 -\frac{B_1 B_3 c_2 \Sigma ^2}{\rho ^2 \rho _2} \\[2mm]
 B_1 c_1 & B_2 c_1 & B_3 c_1 & B_1 c_2 & B_2 c_2 & B_3 c_2 & v_1 & 0 & 0 & \beta ^2 & \frac{\rho _1 
 \rho _2 \BB\cdot\ww}{\rho ^2} & \frac{c_1 c_2 \BB\cdot\ww}{\rho _1} & -\frac{c_1 c_2 
 \BB\cdot\ww}{\rho _2} 
 \\[2mm]
 0 & 0 & 0 & 0 & 0 & 0 & 0 & v_1 & 0 & 0 & 0 & 0 & 0 \\[2mm]
 0 & 0 & 0 & 0 & 0 & 0 & 0 & 0 & v_1 & 0 & 0 & 0 & 0 \\[2mm]
 0 & 0 & 0 & 0 & 0 & 0 & \frac{\Sigma ^2}{\rho ^2} & 0 & 0 & v_1 & -\frac{B_1 \rho_{1,2}  \Sigma 
 ^2}{\rho ^3} & -\frac{B_1 c_1 \Sigma ^2}{\rho ^2 \rho _1} & -\frac{B_1 c_2 \Sigma 
 ^2}{\rho ^2 \rho _2} \\[2mm]
 0 & 0 & 0 & 0 & 0 & 0 & 0 & 0 & 0 & 0 & v_1 & 0 & 0 \\[2mm]
 \rho _1 & 0 & 0 & 0 & 0 & 0 & 0 & 0 & 0 & 0 & \frac{c_2 \rho _1^2 w_1}{c_1 \rho } & v_{1,1} & 0 
 \\[2mm]
 0 & 0 & 0 & \rho _2 & 0 & 0 & 0 & 0 & 0 & 0 & \frac{c_1 \rho _2^2 w_1}{c_2 \rho } & 0 & v_{2,1} \\
\end{array}
\right).
\]
Here, $\rho_{1,2}=\rho_1-\rho_2$, $p_{1,2} = p_1-p_2$, $C_a^2 = \frac{\pd p_a}{\pd \rho_a}$, 
$a=1,2$.

\begin{figure}
	\centering
	\subfloat{\includegraphics[scale=0.35]{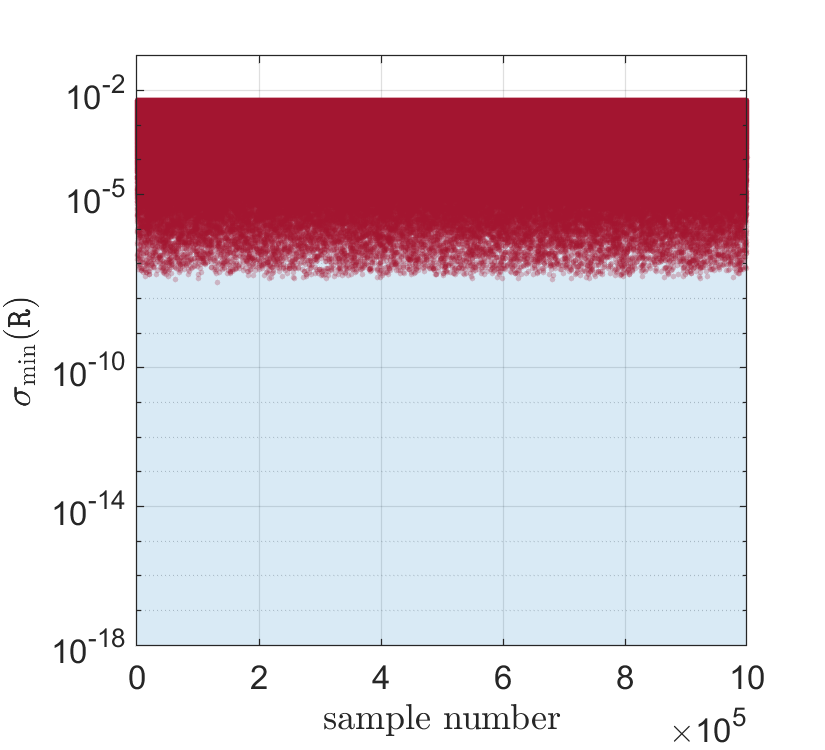}}
	\caption[Put some things in here]{The smallest singular value (red dots) of the matrix 
	$\mathtt{R}(\bm{P})$ of right eigenvectors of $\mathtt{A}(\bm{P})$.}
	\label{fig.7}
\end{figure}

To prove that system \eqref{eqn.quasilin} is hyperbolic, one needs to demonstrate that all 
eigenvalues of matrix $\mathtt{A}(\bm{P})$ are real and right eigenvectors form a basis. 
Because an analytical expression of eigenvectors and eigenvalues are not available for general 
state vector $\bm{P}$, to get at least some evidences in favor of hyperbolicity of 
\eqref{eqn.quasilin}, we performed a numerical study of the eigenvalue decomposition of 
$\mathtt{A}(\bm{P})$ in a domain of the state space near the mechanical and thermodynamical 
equilibrium. Thus, we generated sample vectors $\bm{P}$ with randomly distributed components in the 
following intervals: $ \alpha\in[0.001,0.999]$, $\rho_a\in[0.9\cdot\rho_{a0},1.1\cdot\rho_{a0}]$, 
$v_{a,k}\in[-10,10]$, $B_k\in[-1,1]$. Other material parameters such as the reference mass 
densities 
$\rho_{a0}$, the reference sound speeds $c_{a0}$, the capillarity modulus $\Sigma=100$, the 
microinertia modulus $\beta=1$, and the 
equations of state were taken as in Section \ref{sec.bubble}. We then computed the eigenvalues and 
the matrix of right eigenvectors $\mathtt{R}(\bm{P})$ of $\mathtt{A}(\bm{P})$ numerically using the 
Matlab 
software \cite{MATLAB} and the \texttt{eig} 
function with the options \texttt{eig($\mathtt{A}$,$\mathtt{I}$,'qz')}, where $\mathtt{I}$ is the 
identity matrix of the same size as $\mathtt{A}$. We then checked if the matrix 
$\mathtt{R}(\bm{P})$ 
is singular by computing its singular value decomposition. Fig.\,\ref{fig.7} shows the smallest 
singular value $\sigma_{\min}(\mathtt{R})$ (red dots) for $10^6$ sample state vectors $\bm{P}$. As 
it can be seen the smallest singular value is well separated (blue area) from 0 indicating that the 
matrix 
$\mathtt{R}$ is non-singular for the tested sample vectors $\bm{P}$. During this and other checks 
the numerically computed eigenvalues were always real.

Finally, remark that the analytical expressions of the eigenvalues and vectors are available for 
the case 
of stationary medium $\vv_1=\vv_2=0$. The eigenvalues are 
completely decoupled and read
\begin{equation}
	\lambda_{1,2} = \pm C_1^2,
	\qquad
	\lambda_{3,4} = \pm C_2^2,
	\qquad
	\lambda_{5,6} = \pm \frac{\Sigma}{\rho}\sqrt{\Vert \BB \Vert^2 - B_1^2 + \beta^2}, 
	\qquad
	\lambda_{7,\ldots,13} = 0,
\end{equation}
while the eigenvectors are linearly independent in this case.

\section{Relaxation limit system via formal asymptotic analysis}\label{sec.Asympt}

System \eqref{MasterSysSF} contains a set of dissipative relaxation source terms with 
corresponding relaxation parameters. For some types of two-phase flows, such as 
dispersed flows (bubbly fluids, two-phase flows in porous media), a simplified model can be 
considered when assuming 
very fast 
(instantaneous) relaxation of the thermodynamic forces to the equilibrium. For example, the 
assumption of instantaneous pressure relaxation seems reasonable from a physical point of view, 
since the 
transition of the medium to an equilibrium state is determined by a few passages of fast pressure 
waves 
at
the scale of dispersed inclusions. In many ways, the narrow mixing zone of a diffuse interface 
separating two fluids can be also considered as a dispersed zone, and therefore the above argument 
also applies.
 
In this section, we present a derivation of the so-called relaxation limit of equations 
\eqref{MasterSysSF} assuming instantaneous relaxation of all dissipative processes, namely 
relaxation of the microinertia thermodynamic force $E_D$, pressure $E_\alpha$, and relative 
velocity 
$E_{w_k}$.

First, considering $\ceps = \delta^2/ \lambda \ll 1$ as a small parameter, we derive a relaxation 
limit for \eqref{MasterSysSF} assuming instantaneous relaxation of the thermodynamic force $E_D$ in 
Sec.\,\ref{sec.relax.Limit.D}.
By these means, it will be demonstrated that 
in leading-orders in $\ceps$, the Young-Laplace law is fulfilled locally, in the mixture 
elements.  Then, in Sec.\ref{sec.relax.limit.all}, we consider relaxation of the remaining 
thermodynamic forces $E_\alpha$ and $E_{w_k}$, and derive the final isentropic single-velocity 
approximation of \eqref{MasterSysSF}.

\subsection{Relaxation limit of \eqref{MasterSysSF}}\label{sec.relax.Limit.D}

For our purposes, it is enough to consider only two equations for $\alpha$ and $D$:
\begin{subequations}\label{eqn.approx}
	\begin{align}
		\rho \md{\alpha} &= -\frac{1}{\delta}\rho E_D,\label{eqn.exp.a}
		\\[2mm]
		\rho \md{D} + \frac{\pd\left( \rho E_{B_k} \right)}{\pd x_k} 
		&=
		\frac{1}{\delta} \rho E_\alpha - \frac{1}{\ceps}\rho E_D,\label{eqn.approx.D}
	\end{align}
\end{subequations}
where we have used the material time derivative $\rmd/\rmdt = \pd/\pd t + v_k \pd/\pd x_k$ to 
rewrite the equations in a more compact form.

We then expand $D$ and the thermodynamic force $E_D$ 
\begin{subequations}\label{eqn.expansion}
	\begin{align}
		D & = D_0 + \epsnd D_1 + \epsnd^2 D_2+\ldots,\\[2mm]
		E_{D} &=E_{D,0} + \epsnd E_{D,1} + \epsnd^2 E_{D,2} +\ldots,
	\end{align}
\end{subequations}
in powers of the small non-dimensional parameter $ \epsnd = 
\ceps/\ceps_0 $, where $\ceps_0$ is a time scale. Then, plugging this into \eqref{eqn.approx.D}
\begin{equation}
	\rho \md{}\left( D_0 + \epsnd D_1 + \calO(\epsnd^2) \right)  + \frac{\pd\left( \rho E_{B_k} 
	\right)}{\pd x_k} =
	\frac{1}{\delta} \rho E_\alpha - \frac{1}{\ceps_0 \epsnd}\rho \left( E_{D,0} + \epsnd E_{D,1} + 
	\calO(\epsnd^2)
	\right),
\end{equation}
we obtain that 
\begin{equation}
	E_{D,0} = \calO(\epsnd).
\end{equation}
Hence, assuming $E_{D,0} = 0$ and $D_0 = E_{D,0}/\beta^2=0$, we obtain the following approximate 
equation on 
$E_{D,1}$ 
\begin{equation}
		\rho \md{}\left( \epsnd D_1 + \calO(\epsnd^2) \right)  + \frac{\pd\left( \rho E_{B_k} 
		\right)}{\pd x_k} =
		\frac{1}{\delta} \rho E_\alpha - \frac{1}{\ceps_0 \epsnd}\rho \left( \epsnd 
		E_{D,1} + 
		\calO(\epsnd^2)
		\right),
\end{equation}
from which, we deduce that
\begin{equation}
	E_{D,1} = \ceps_0 \frac{1}{\delta}E_\alpha - \ceps_0 \frac{1}{\rho} \frac{\pd}{\pd x_k} \left( 
	\rho E_{B_k} \right) + \calO(\epsnd).
\end{equation}

On the other hand, the equation for $\alpha$ can be approximated as
\begin{equation}
	\delta \rho \md{\alpha} = -\rho \epsnd E_{D,1} + \calO(\epsnd^2),
\end{equation}
and then, using the expression of $E_{D,1}$, the relaxation limit equation for volume fraction reads
\begin{equation}\label{eqn.approx.a}
	\rho \md{\alpha} = -\frac{1}{\lambda}\rho E_\alpha + \frac{1}{\lambda}\delta \frac{\pd}{\pd 
	x_k} \left( \rho 
	E_{B_k} \right) + \calO(\epsnd^2).
\end{equation}
From this equation, we can conclude that, in contrast to the master system \eqref{MasterSys}, the 
time evolution of the volume fraction is governed not only by
the pressure relaxation $\lambda^{-1}\rho E_\alpha = \lambda^{-1}(p_2-p_1)$, but also by the 
curvature of the diffuse interface. To see the latter, one needs to use the explicit expression of 
$E_{B_k} 
= \SigmaMod B_k/\rho^2$, the definition $ B_k = \delta \frac{\pd \alpha}{\pd x_k}$, and the assumption that 
the surface tension coefficient $\sigma=\delta \SigmaMod/\rho $ is constant, $\sigma=const$. After 
this, equation \eqref{eqn.approx.a} becomes (we omit the terms $\calO(\epsnd^2)$)
\begin{equation}\label{eqn.approx.a2}
	\rho \md{\alpha} = -\frac{1}{\lambda}\rho E_\alpha + \frac{1}{\lambda}\delta \sigma 
	\frac{\pd}{\pd x_k} \left( 
	\frac{\pd \alpha}{\pd x_k}  \right),
\end{equation}
where we used definition \eqref{eqn.Sigma.scaling1} of the surface tension coefficient $\sigma$.

In particular, for stationary flows ($\md{\alpha}=0$), the later equation reduces to
\begin{equation}\label{eqn.YL.diffuse1}
    \delta \sigma \frac{\pd}{\pd x_k} \left( \frac{\pd \alpha}{\pd x_k} 
    \right)  = p_2 - p_1.
\end{equation}

To obtain the Young-Laplace law
\begin{equation}\label{eqn.YL.diffuse2}
    \sigma \nabla\cdot \left( \frac{\nabla\alpha}{\Vert \nabla\alpha\ \Vert }\right) 
    = p_2 - p_1,
\end{equation}
from \eqref{eqn.YL.diffuse1}, the parameter $\delta = \Vert \del \alpha \Vert^{-1}$ must appear 
under 
the outer derivative $\pd/\pd x_k$, but $\delta$ is not constant and this cannot be done without 
altering \eqref{eqn.YL.diffuse1}. However, 
we shall show in Sec.\,\ref{sec.bubble}, that in practical computations, $\delta$ can be replaced 
by an a priory computed constant (see \eqref{eqn.delta0}), so that in fact \eqref{eqn.YL.diffuse1} 
approximates the 
Young-Laplace law \eqref{eqn.YL.diffuse2}.

Finally, we note that the relaxation limit of \eqref{MasterSysSF} for $\ceps\to0$ is system 
\eqref{MasterSysSF} without equation for $D$ and in which the equations for $\alpha$ and $B_k$ are 
replaced by the following:
\begin{subequations}\label{eqn.relax.limit.parab}
	\begin{align}
		\rho \md{\alpha} &-\frac{\delta}{\lambda} \frac{\pd}{\pd 
					x_k} \left( \rho 
					E_{B_k} \right)= -\frac{1}{\lambda}\rho E_\alpha,   \\[2mm]
		\md{B_k} &+ B_j \frac{\pd v_j}{\pd x_k} + \frac{\pd}{\pd x_k}\left( 
		\frac{\delta}{\lambda}E_\alpha \right) - \frac{\pd}{\pd x_k} \left( 
		\frac{\delta^2}{\lambda\rho} \frac{\pd \rho E_{B_j}}{\pd x_j}\right) = 0,
	\end{align}
\end{subequations}
where the equation for $B_k$ is a second order parabolic equation, that can be obtained by the same 
means as \eqref{eqn.approx.a}.

One could note the antisymmetric structure (opposite signs) of the constitutive fluxes in 
\eqref{eqn.relax.limit.parab} involving the
thermodynamic forces $E_\alpha$ and $E_{B_k}$. This usually results in complex eigenvalues, and
subsequently in the loss  
of hyperbolicity of the first-order differential operator (without dissipative parabolic term 
in the second equation of \eqref{eqn.relax.limit.parab}). Therefore, in the next section we 
consider another relaxation limit of \eqref{MasterSysSF}, in which we couple the result of this 
section with the single velocity approximation ($\vv_1=\vv_2$).


\subsection{Relaxation limit of the single velocity isentropic model}
\label{sec.relax.limit.all}
In this section, on top of the previous result, we derive reduced single-velocity isentropic 
equations obtained as a relaxation 
limit of \eqref{MasterSysSF} when $E_\alpha$ and $E_{w_k}$ are instantaneously set to their 
equilibrium values. The latter can be obtained by simply assuming the relative velocity to be 
zero $\ww=\vv_1-\vv_2 = 0$, since $E_{\ww} = c(1-c) \ww$.
Thus, the single-velocity approximation system is a consequence of \eqref{MasterSysSF} under the 
assumption
$w_k=0$, that can be written in the following form 
\begin{subequations}\label{MasterSysSinglVel}
	\begin{align} 
	 &\frac{\pd \rho_1 \alpha_1}{\pd t} +
	\frac{\pd \rho_1 \alpha_1 v_{k}}{\pd x_{k}}= 0, \label{mass1}
		\\[2mm]
     &\frac{\pd \rho_2 \alpha_2}{\pd t} +
	\frac{\pd \rho_2 \alpha_2 v_{k}}{\pd x_{k}}= 0, \label{mass2}
		\\[2mm]
	 &\frac{\pd \rho v_i }{\pd t}+
	\frac{\pd \left(\rho v_i v_k + \rho^2 E_\rho \delta_{ki} + 
				\rho B_i E_{B_k} \right)}{\pd x_{k}}= 0, 
		\\[2mm]
	&\frac{\pd \rho\alpha }{\pd t} +
	 \frac{\pd \rho\alpha v_{k}}{\pd x_{k}} = - \frac{1}{\delta}\rho E_D, 
	 \\[2mm]
	& \frac{\pd \rho D}{\pd t} + \frac{\pd (\rho D v_k+\rho E_{B_k})}{\pd x_{k}} 
	=\frac{1}{\delta} \rho E_\alpha - \frac{1}{\varepsilon} \rho E_D ,
		\\[2mm]
	&
	\frac{\pd B_k}{\pd t} + \frac{\pd (B_l v_l+E_D)}{\pd x_{k}}+
	 v_{l} \left( \frac{\pd B_k}{\pd x_{l}}-\frac{\pd B_l}{\pd x_{k}} \right)  =0.
	\end{align}
\end{subequations}

In the previous section, we derived the asymptotic equation \eqref{eqn.approx.a} for the volume 
fraction assuming small relaxation time for the microinertia field $D$. In turn, further assuming 
that the 
relaxation parameter $\lambda$ is sufficiently small so that the time variation of the volume 
fraction is 
small in comparison with the relaxation rate $\lambda^{-1}$, equation \eqref{eqn.approx.a} can be 
approximated as
\begin{equation} \label{rho12Eqn}
p_2(\rho_2)-p_1(\rho_1)={\delta}\frac{\pd \rho E_{B_k}}{\pd x_{k}}.
\end{equation}

From \eqref{rho12Eqn}, we immediately obtain two relations for derivatives of phase mass densities
\begin{equation} \label{denrelations}
C^2_2\frac{\pd \rho_2}{\pd t} -  
C^2_1\frac{\pd \rho_1}{\pd t} = \frac{\pd }{\pd t}\left( {\delta}\frac{\pd \rho E_{B_k}}{\pd 
x_{k}} \right), \qquad
C^2_2\frac{\pd \rho_2}{\pd x_j} -  
C^2_1\frac{\pd \rho_1}{\pd x_j} = \frac{\pd }{\pd x_j} \left( {\delta}\frac{\pd \rho E_{B_k}}{\pd 
x_{k}} \right),
\end{equation}
where $C_a^2 = \pd p_a/\pd \rho_a$, $a=1,2$ are the phase adiabatic sound speeds.

Now, using relations \eqref{denrelations} and phase mass conservation equations \eqref{mass1}, 
\eqref{mass2},  one can obtain the following equation for $\alpha_1$:
\begin{equation}\label{eqforalpha}
    \md{\alpha_1}+
    \frac{\alpha_1\alpha_2(K_1-K_2)}{\alpha_2K_1+\alpha_1K_2}
    \frac{\pd v_k}{\pd x_k}-
    \frac{\alpha_1\alpha_2}{\alpha_2K_1+\alpha_1K_2}
    \md{}\left( {\delta}\frac{\pd \rho E_{B_k}}{\pd x_{k}} \right)=0, 
\end{equation}
where $K_a=\rho_a C_a^2$ are the phase bulk moduli.

If we recall the definition $B_k=\delta \partial \alpha_1/\partial x_k$,
 then it is clear that \eqref{eqforalpha} is a third-order partial differential equation for 
 $\alpha=\alpha_1$.
Finally, we can also formulate a closed relaxation limit system of the single velocity isentropic 
approximation of \eqref{MasterSysSF} for the variables  
 $\alpha_1$, $\rho_1$, $\rho_2$, and $v_k$:
\begin{subequations}\label{ResultSysSinglVel}
	\begin{align} 
      & \frac{\pd \alpha_1}{\pd t}+v_k \frac{\pd \alpha_1}{\pd x_k}+
    \frac{\alpha_1\alpha_2(K_1-K_2)}{\alpha_2K_1+\alpha_1K_2}
    \frac{\pd v_k}{\pd x_k}-
    \frac{\alpha_1\alpha_2}{\alpha_2K_1+\alpha_1K_2}
    \md{}\left( \delta \sigma \frac{\pd^2 \alpha_1}{\pd x_k \pd 
    x_{k}} \right)=0, 
    \\[2mm]
	 &\frac{\pd \rho_1 \alpha_1}{\pd t} +
	\frac{\pd \rho_1 \alpha_1 v_{k}}{\pd x_{k}}= 0, 
		\\[2mm]
  &\frac{\pd \rho_2 \alpha_2}{\pd t} +
	\frac{\pd \rho_2 \alpha_2 v_{k}}{\pd x_{k}}= 0,
		\\[2mm]
	 &\frac{\pd \rho v_i }{\pd t}+
	\frac{\pd }{\pd x_{k}} \left(\rho v_i v_k +  \left( p_1 + \alpha_2  
	\left( \delta \sigma \frac{\pd^2\alpha_1 }{\pd x_j \pd 
	x_j}\right) \right) \delta_{ki}
	+
	\hat{T}_{ki}
		\right)= 0, 
	\end{align}
\end{subequations}
where $\hat{T}_{ki} = - \frac{\SigmaMod}{\rho} \left( B_j B_j \delta_{ki} - B_k B_i \right)$ 
is the surface tension stress tensor which should also be expressed in terms of the gradients $ 
\pd\alpha_1/\pd x_k$ using the definition of $B_k$.

Note that if surface tension is neglected, then terms with higher derivatives of $\alpha_1$ and the 
tensor $\hat{T}_{ki}$ should be excluded from \eqref{ResultSysSinglVel}.
In this case, \eqref{ResultSysSinglVel} transforms to the well-known five-equation two-phase model 
of Kapila \cite{MurroneGuillard2005}.

The presence of third-order derivatives in \eqref{ResultSysSinglVel} and the corresponding 
dispersion effects of the model can lead to non-standard behavior of waves and this will be the 
subject of further research.

\section{Stationary solution of a spherical bubble}
\label{sec.bubble}

In the previous section, we demonstrated that in the limit $\delta\to 0$ and $\lambda\to 0$, in the 
leading-order terms, the pressure difference inside a mixture element fulfills the relation 
\eqref{eqn.YL.diffuse1}, that in turn resembles the Young-Laplace law \eqref{eqn.YL.diffuse2}. The 
formal obstacle to identify the two relations \eqref{eqn.YL.diffuse1} and \eqref{eqn.YL.diffuse2} 
is the parameter $\delta = \Vert \del\alpha \Vert^{-1}$ that appears outside of the divergence 
operator. In 
what follows, we shall demonstrate that, in fact, $\delta$ can be chosen as a constant so that the  
Young-Laplace law holds on macroscopic diffuse interfaces between two immiscible fluids in an 
approximation sense. 

We shall search for a spherically symmetric stationary solution to system \eqref{MasterSysSF} 
representing a bubble or droplet of a radius $R$. Thus, we assume that $\vv=0$ and $\ww=0$.
We also assume that the temperature variations are negligible, that means that we can exclude the 
entropy 
from consideration.  
All unknown scalar functions can be parameterized as
\begin{equation}\label{eqn.hat.a}
	\alpha(t,\xx) = \tilde{\alpha}(r),
	\qquad 
	\rho(t,\xx) = \tilde{\rho}(r),
	\qquad
	D(t,\xx) = \tilde{D}(r),
	\qquad
	r = \Vert\xx\Vert
\end{equation}
and vector fields as
\begin{equation}\label{eqn.hat.B}
	B_k(t,\xx) = \tilde{B}(r) \frac{x_k}{r}.
\end{equation}
Moreover, we prescribe the distribution of $\tilde{\alpha}$ in the diffuse interface between the 
fluids in 
the form
\begin{equation}\label{eqn.tanh}
	\tilde{\alpha}(r) = \frac{1}{2} \left( 1 - \tanh\left( \frac{\pi (r-R)}{h} \right) \right), 
\end{equation}
where $h$ is the thickness of the diffuse interface, and we will look for a solution that 
satisfy 
\eqref{eqn.tanh}. This also fixes $\tilde{B}(r)$ as
\begin{equation}\label{eqn.B.amp}
	\tilde{B}(r) =\delta \tilde{\alpha}'(r)
\end{equation}
due to the definition \eqref{B&D} of the vector field $B_k$. In the following, we shall omit the 
tilde sign 
``$\tilde{\phantom{\alpha}}$'' above the unknowns for simplicity of notations. The reader should 
keep in mind that all the state variables are functions of the single variable $r$.

It is sufficient to consider the momentum equations and equation on $D$ that, for a steady state, 
reduce to 
\begin{subequations}\label{StatSys}
	\begin{align} 
		&\frac{\pd (\rho^2 E_\rho \delta_{ki} + 
		\rho B_i E_{B_k})}{\pd x_{k}}= 0, \quad i=1,2,3, 
		\label{StatMomentum}\\
		& \frac{\pd (\rho E_{B_k})}{\pd x_{k}} =\frac{1}{\delta} \rho E_\alpha.
		\label{StatD}
	\end{align}
\end{subequations}

After using the spherical symmetry assumptions \eqref{eqn.hat.B}, definition of $B(r)$ 
\eqref{eqn.B.amp}, and the relation $\sigma = \delta \SigmaMod/\rho$ \eqref{eqn.Sigma.scaling1}, 
these two equations reduce to an ordinary differential equation and an algebraic equation (because 
$\alpha(r)$ is given by \eqref{eqn.tanh})
\begin{subequations}
	\begin{align}
		p'(r)  		&=-\frac{2 \sigma \delta(r) \alpha'(r)^2}{r} \equiv F(r),\label{eqn.p'}\\
		p_2(r) - p_1(r) 	&= \sigma \delta(r) \left( \frac{d-1}{r} \alpha'(r) + 
		\alpha''(r)\right) \equiv G(r),
	\end{align}
\end{subequations}
where $d$ is the space dimension ($d=3$ in this example).

The first equation can be integrated to get the mixture pressure $p(r) = \alpha(r) p_1(r) + 
(1-\alpha(r))p_2(r)$
\begin{equation}\label{eqn.bubble.p}
	p(r) = p_{\textrm{atm}} - \int\limits_{r}^{\infty} F(\zeta) \rmd \zeta,
	\qquad
	p_{\textrm{atm}} = p(\infty),
\end{equation}
after which, the phase pressures can be found as
\begin{equation}\label{eqn.bubble.p12}
	p_1(r) = p(r) - ( 1 - \alpha(r) ) G(r),
	\qquad
	p_2(r) = p(r) + \alpha(r)   G(r).
\end{equation}

In Section\,\ref{sec.closure}, we concluded that to recover the conventional capillary stress tensor \eqref{eqn.T.classic}, one should take $\delta(r) = \Vert \del \alpha \Vert^{-1}$. This, however, is 
not convenient from the computational view point, because $\del\alpha$ is not a state variable of the 
system \eqref{MasterSysSF} and cannot be easily evaluated. Therefore, a practical procedure has to 
be invented in order to provide a reasonable estimate of $\delta$ \textit{a priori}. 
In this paper, we suggest replacing $\delta(r)$ by the following constant 
$\delta_0=\text{const}$
\begin{equation}\label{eqn.delta0}
	\delta_0(h) = \int\limits_{R-h}^{R+h} r^{-1}\alpha'(r) \rmd r
	\cdot \left( \int\limits_{R-h}^{R+h} 
	r^{-1}\alpha'(r)^2 \rmd r \right)^{-1},
\end{equation}
which is choosing in such a way to obtain the Young-Laplace formula $\Delta p=-2\sigma/R$ in  \eqref{eqn.p'}. Exactly this $\delta_0$ is used instead 
of 
$\delta=\Vert \del\alpha \Vert^{-1}$ in the numerical results below.

\begin{figure}
	\centering
	\subfloat{\includegraphics[scale=0.65]{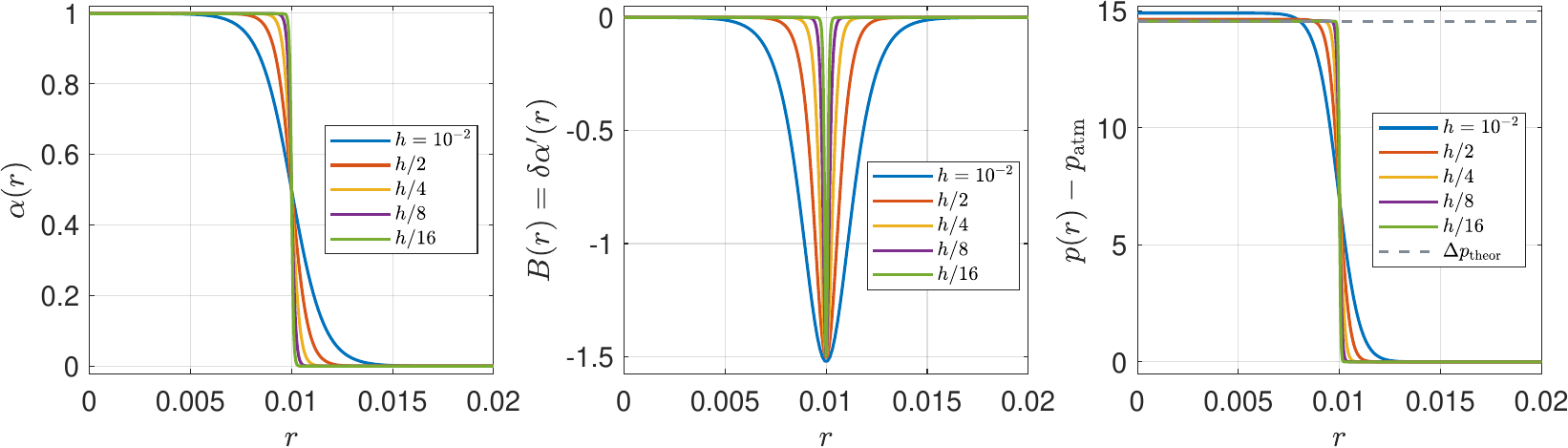}}
	\caption[Put some things in here]{Diffuse interface profile for the stationary bubble of radius 
	$R = 0.01$, $\sigma = 0.0728$ for which the theoretical pressure jump $\Delta 
	p_{\textrm{theor}} = 
	(d-1) \sigma/R = 14.56$. Left: volume fraction for various interface width $h$. Right: mixture 
	pressure $p(r) = \alpha p_1 + (1-\alpha)p_2$}
	\label{fig.YL.width}
\end{figure}

Typical solutions to \eqref{eqn.bubble.p} for $R=0.01$, $\sigma = 0.0728$, and various interface 
widths 
$h$ are depicted in 
Fig.\,\ref{fig.YL.width}. In particular, one can see that as long as $h\to0$ the pressure jump 
across 
the interface converges to the theoretical one $\Delta p_{\textrm{theor}} = (d-1) \sigma/R=14.56$ 
given 
by the Young-Laplace law. One should bear in mind that for every curve in Fig.\,\ref{fig.YL.width}, 
$\delta_0(h)$ is different since it depends on $h$.

\begin{figure}
	\centering
	\subfloat{\includegraphics[scale=0.65]{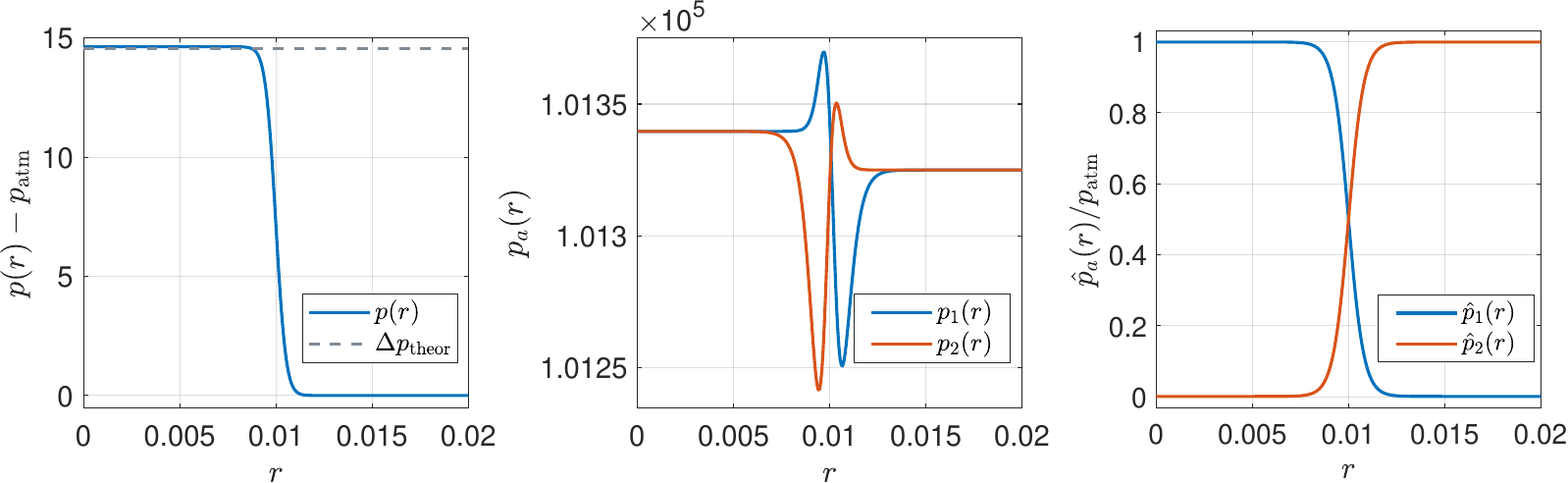}}
	\caption[Put some things in here]{Mixture pressure $p(r)$ (left), phase pressures $p_a(r)$ 
	(middle), 
	and partial pressures $\hat{p}_a(r) = \alpha_a p_a$ (right), for a stationary bubble with 
	$R=0.01$, $\sigma = 0.0728$, and $h=0.005$.}
	\label{fig.YL.p}
\end{figure}

Fig.\,\ref{fig.YL.p} depicts the phase pressure $p_a(r)$ and partial 
pressure $\hat{p}_a(r) = \alpha_a(r)p_a(r)$ variations across the interface. It demonstrates the 
main feature of our model, that is, the pressures $p_1$ and $p_2$ of the mixture constituents can 
be different, and 
that the presence of the interface curvature prevent $p_1$ and $p_2$ from relaxing to a common 
value as it is assumed in the single-pressure models, e.g. 
\cite{Brackbill1992,Saurel2005,Schmidmayer2017,Chiocchetti2021}.

For this particular solution, one does not need to specify the fluid equations of state 
$e_a(\rho_a,S)$ in \eqref{eqn.E1}, however, it would be interesting to take some particular 
internal energies $e_a(\rho_a,S)$ and to look at the density variation across the interface. Thus, 
we shall assume that the fluid with index $a=1$ (in the center of the domain, left from the 
interface in the figures) is an ideal gas given by the ideal gas equation of 
state, while the fluid $a=2$ is a liquid parameterized by the stiffened-gas equation of state
\begin{equation}\label{eqn.eos}
	e_1(\rho_1,S) = \frac{c_{10}^2}{\gamma_1(\gamma_1-1)} \left( \frac{\rho_1}{\rho_{10}} 
	\right)^{\gamma_1-1} e^{S/c_{V,1}},
	\qquad
	e_2(\rho_2,S) = \frac{c_{20}^2 }{\gamma_2(\gamma_2-1)} \left( 
	\frac{\rho_2}{\rho_{20}} 
	\right)^{\gamma_2-1} e^{S/c_{V,2}} + 
	\frac{\rho_{20}c_{20}^2 - \gamma_2 p_{20}}{\gamma_2 \rho_2},
\end{equation}
where $c_{V,a}$, $a=1,2$ are the specific heats at constant volume, $\gamma_a$ are the ratio of the 
specific heats, $c_{a0}$ are the reference sound speeds, $\rho_{a0}$ are the reference 
phase mass densities, and 
$p_{20}$ is the reference pressure of the liquid phase. 

\begin{figure}
	\centering
	\subfloat{\includegraphics[scale=0.65]{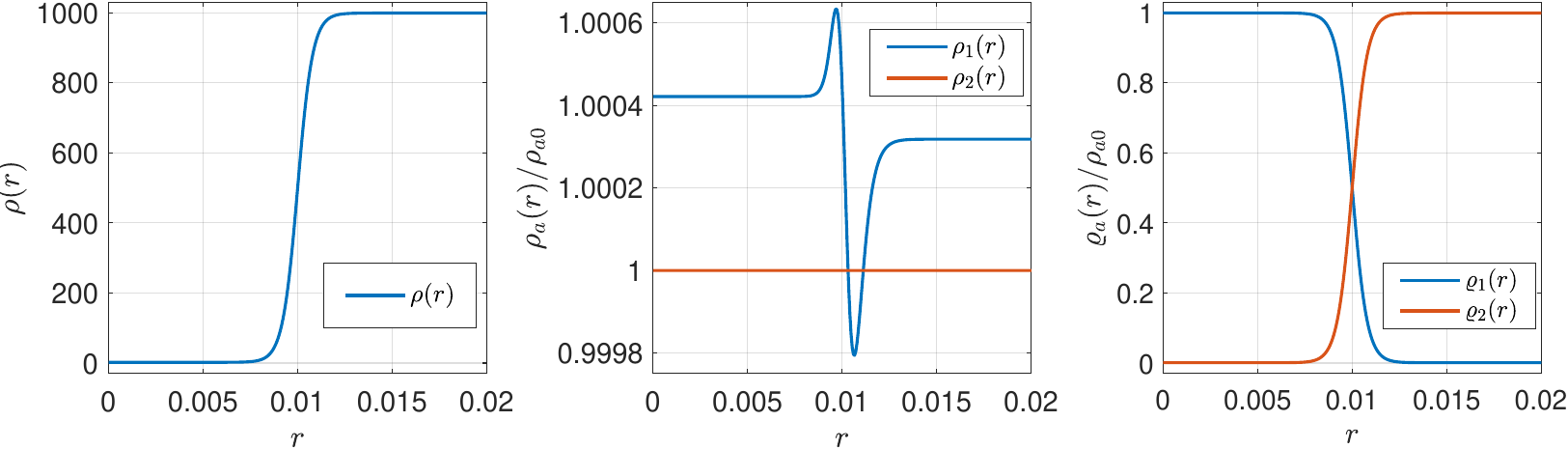}}
	\caption[Put some things in here]{Mixture mass density $\rho(r) = \alpha \rho_1 + 
	(1-\alpha)\rho_2$  
	(left), phase mass densities $\rho_a(r)$ (middle), 
		and apparent densities $\varrho_a(r) = \alpha_a \rho_a$ (right), for a stationary bubble 
		with 
		$R=0.01$, $\sigma = 0.0728$, and $h=0.005$. The ideal gas and stiffened-gas equations of 
		state 
		were 
		used to obtain fluid densities $\rho_a$ from pressures $p_a$.}
	\label{fig.YL.rho}
\end{figure}

Fig.\,\ref{fig.YL.rho} shows typical density profiles recovered from the pressure profiles using 
equations of state \eqref{eqn.eos}. The following values were used for the gas phase 
$\gamma_1=1.4$, $ c_{10} =343.03$, $c_{V,1}=717.2$, 
$S=0$ that gives the equilibrium density of the gas $\rho_{10} = 1.205$. The liquid parameters 
were taken as follows $\gamma_2=1.949437$, $c_{V,2}=4150$, 
$p_{20 } = p_{\textrm{atm}} = 101325$, $c_{20} = 1500$, $\rho_{20}=1000$. Computing the phase 
pressures as
$p_a = \rho_a^2 \frac{\pd e_a}{\pd \rho_a}$ and inverting it with respect to $\rho_a$, we can plot 
the density profiles as depicted in Fig.\,\ref{fig.YL.rho}. Note that the liquid density 
$\rho_2(r)$ is not exactly constant as it may seem from Fig.\,\ref{fig.YL.rho} (middle) but its 
perturbations are of the order $10^{-8}$.

\section{Dispersion relations}\label{sec.dispers}

In this section, we perform a linear stability analysis of \eqref{MasterSysSF} by considering a 
particular solution in the form of a plane wave
\begin{equation}
	\bm{Q}(t,\bm{x}) = \bm{Q}_0 e^{i (\omega t- \bm{k} \cdot \bm{x})},
\end{equation}
where $i=\sqrt{-1}$ is the imaginary unit, $\omega = 2\pi f$ is the real angular frequency, $f$ is 
the frequency, $\bm{k}$ is the complex wave vector. It is sufficient to restrict the analysis to 
the 1D case, i.e. $\bm{x}=(x,0,0)$, and $\bm{k} = (k,0,0)$,
and moreover to the genuinely 1D case, i.e. we set 
$(v_{a,1},v_{a,2},v_{a,3}) = (v_{a},0,0)$, $(B_1,B_2,B_3) 
= (B,0,0)$. 

To derive the required PDE system it is necessary to use relations between mixture and individual 
phases
variables of state. Then, after a cumbersome but standard procedure, one can derive equations for 
the state variables of the phases and linearize these equations near the equilibrium state
$v_a=0 + v'_{a}$, $\rho_{a} = 
\rho_{a,0}+\rho'_{a}$, $\alpha=\alpha_0+\alpha'$, $D= 0 + D' $, $B=B_0 + B'$. 

The equations for 
perturbations $\bm{Q}'=(v'_1,v'_2,B',D',\alpha',\rho'_1,\rho'_2)$ reads (we omit the prime symbol 
``$\,'\,$'' for the sake of brevity)
\begin{subequations}\label{eqn.linear}
	\begin{align}
		&\frac{\partial v_1}{\partial t} +\frac{p_{1,0}-p_{2,0}}{\rho _0}\frac{\partial \alpha 
		}{\partial x}+\frac{C_1^2}{\rho _{1,0}}\frac{\partial \rho _1}{\partial 
		x}=-\frac{ c_{1,0} c_{2,0}^2}{\chi }\left(v_1-v_2\right),
		\\
		&\frac{\partial v_2}{\partial t}+\frac{p_{1,0}-p_{2,0}}{\rho _0}\frac{\partial \alpha 
		}{\partial x}+\frac{C_2^2}{\rho _{2,0}}\frac{\partial \rho _2}{\partial 
		x}= +\frac{c_{1,0}^2 c_{2,0}}{\chi }\left(v_1-v_2\right),
		\\
		&\frac{\partial  B}{\partial t}+B_0c_{1,0}\frac{\partial 
		v_1}{\partial x}+B_0 c_{2,0}\frac{\partial 
		v_2}{\partial x}+\beta ^2\frac{\partial D}{\partial x}=0,
		\\
		&\frac{\partial D}{\partial t}+\frac{\Sigma ^2}{\rho _0^2}\frac{\partial  B}{\partial 
		x}-\frac{B_0 \Sigma ^2 \left(\rho _{1,0}-\rho _{2,0}\right)}{\rho _0^3}\frac{\partial 
		\alpha }{\partial x}-\frac{\alpha _0 B_0 \Sigma ^2}{\rho _0^3}\frac{\partial \rho 
		_1}{\partial x}-\frac{\left(1-\alpha _0\right) B_0 \Sigma ^2}{\rho _0^3}\frac{\partial \rho 
		_2}{\partial x}=-\frac{\rho _1 C_{1,0}^2-\rho _2 C_{2,0}^2}{\delta  \rho _0}-\frac{ 
		\beta ^2 \lambda }{\delta ^2} D,
		\\
		&\frac{\partial \alpha }{\partial t}=-\frac{\beta ^2}{\delta }D, 
		\\
		&\frac{\partial \rho _1}{\partial t}+\rho _1\frac{\partial v_1}{\partial x}=\frac{ 
		\beta ^2 \rho _{1,0}}{\alpha _0 \delta }D,
		\\
		&\frac{\partial \rho _2}{\partial t}+\rho _2\frac{\partial v_2}{\partial x}=-\frac{ 
		 \beta ^2 \rho _{2,0}}{\left(1-\alpha _0\right) \delta }D,
	\end{align}
\end{subequations}
where  $C_a^2 = \frac{\pd p_a}{\pd \rho_a}$, $a=1,2$

If linear system \eqref{eqn.linear} is written in the matrix notations as 
\begin{equation}
	\bm{Q}_t + \mathtt{A} \bm{Q}_x = \mathtt{S} \bm{Q}
\end{equation}
then the dispersion relations $k(\omega)$ of \eqref{eqn.linear} are given as the roots of the 
polynomial (e.g. see \cite{Ruggeri1992})
\begin{equation}\label{eqn.poly}
	\det\left( \mathtt{I} - \frac{k}{\omega} \mathtt{A} + \frac{i}{\omega} \mathtt{S} \right) = 0,
\end{equation}
with $\mathtt{I}$ being the identity matrix.

The phase $V_\text{ph}(\omega)$ and group $V_\text{g}(\omega)$ 
velocities and the attenuation factor $A(\omega) $ can be computed as 
\begin{equation}
	V_\text{ph} = \frac{\omega}{\text{Re}(k)}, \qquad
	V_\text{g } = \left( \frac{\pd \text{Re}(k)}{\pd \omega} \right)^{-1},
	\qquad
	A(\omega) = -\text{Im}(k).
\end{equation}

\begin{figure}
	\centering
	\subfloat{\includegraphics[scale=0.65]{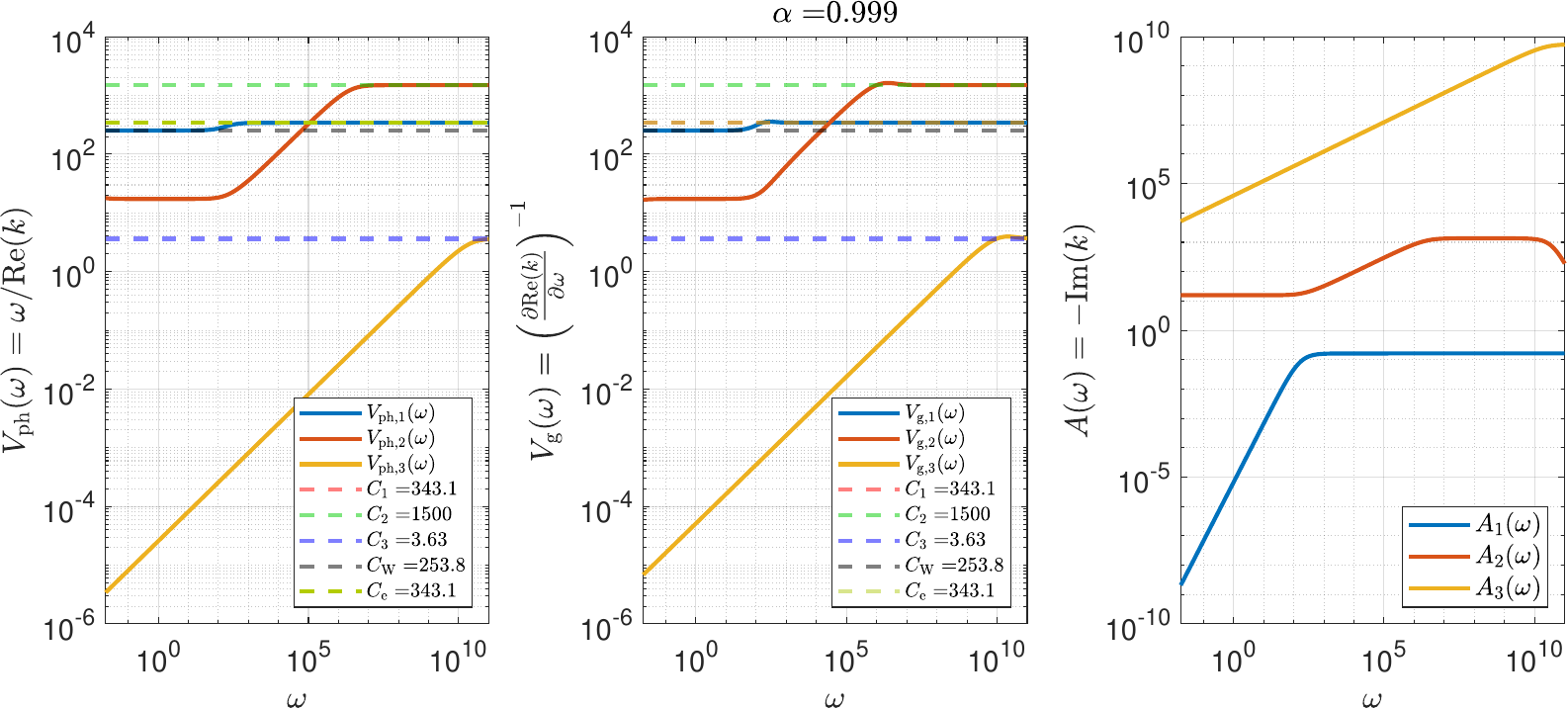}}
	\caption[Put some things in here]{Phase velocities (left), group velocities (middle), and 
	attenuation factor (right) versus the angular frequency $\omega = 2 \pi f$, where $f$ is the 
	frequency. The dashed lines shows the pure gas $C_1=\sqrt{\pd p_1(\rho_1,S)/\pd \rho_1}$ and 
	liquid 
	$C_2=\sqrt{\pd p_2(\rho_2,S)/\pd\rho_2}$ characteristic 
	velocities, 
	equilibrium characteristic speed $C_{\text{e}}$, and Wood's sound speed $C_{\text{W}}$. The 
	mixture parameters are 
	$\alpha=0.999$, $\beta = 1$, $\delta = 0.0025$, $\delta/\lambda = 10^{-8}$. The phase pressures 
	parameters are extracted from the stationary bubble solution at the corresponding value of 
	$\alpha$: $p_1 = p_2 = 1.013395 \cdot 10^{5}$, while other thermodynamic parameters were 
	computed from the equations of state.}
	\label{fig.alpha0.999}
\end{figure}

\begin{figure}
	\centering
	\subfloat{\includegraphics[scale=0.65]{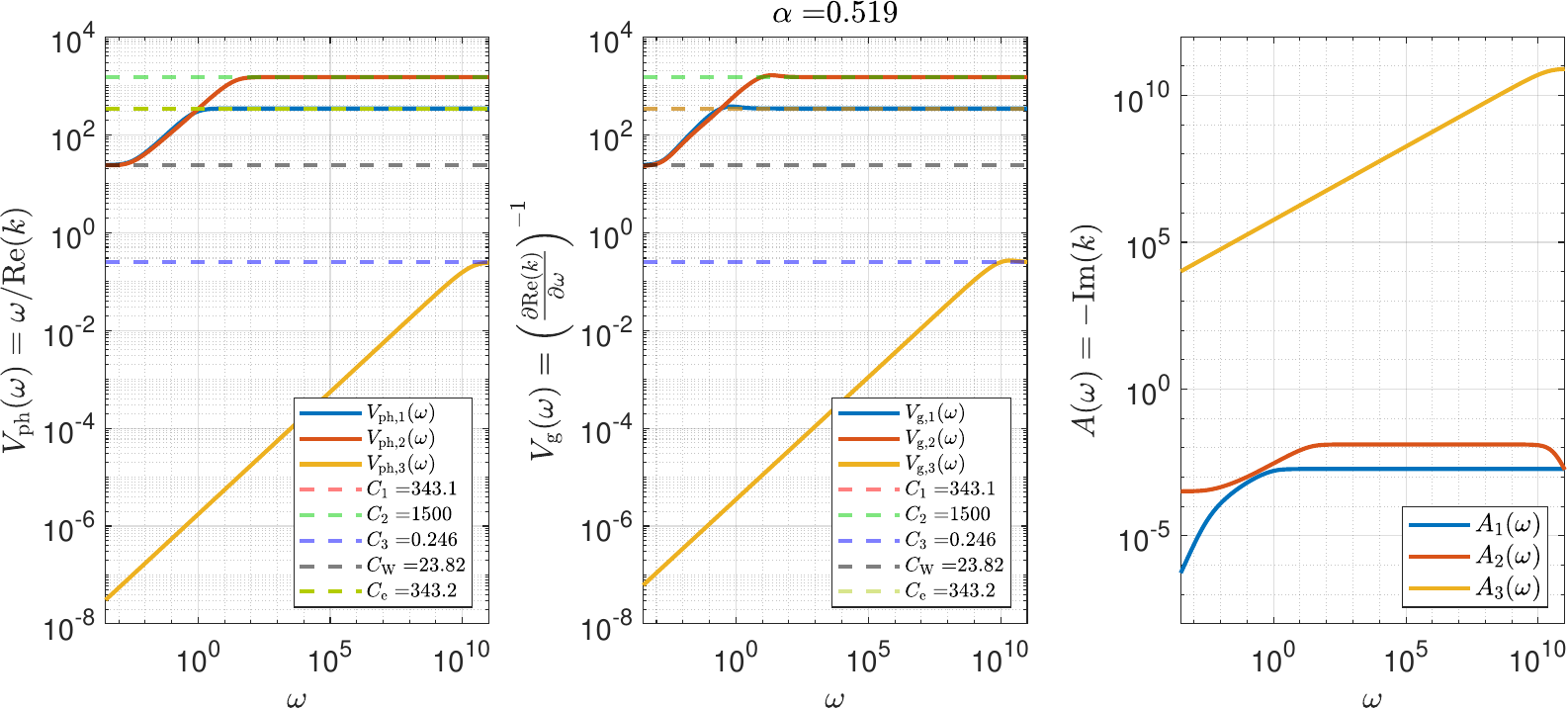}}
	\caption[Put some things in here]{Phase velocities (left), group velocities (middle), and 
		attenuation factor (right) versus the angular frequency $\omega = 2 \pi f$, where $f$ is 
		the 
		frequency. The dashed lines shows the pure gas $C_1=\sqrt{\pd p_1(\rho_1,S)/\pd \rho_1}$ 
		and 
			liquid 
			$C_2=\sqrt{\pd p_2(\rho_2,S)/\pd\rho_2}$ characteristic 
			velocities, 
			equilibrium characteristic speed $C_{\text{e}}$, and Wood's sound speed $C_{\text{W}}$. 
			The mixture parameters are 
		$\alpha=0.519$, $\beta = 1$, $\delta = 0.0025$, $\delta/\lambda = 10^{-8}$. The phase 
		pressures 
		parameters are extracted from the stationary bubble solution at the corresponding value of 
		$\alpha$: $p_1 = 1.013531 \cdot 10^{5}$, $p_2 = 1.013103\cdot10^{5}$, while other 
		thermodynamic parameters were computed from the equations of state.}
	\label{fig.alpha0.519}
\end{figure}

\begin{figure}
	\centering
	\subfloat{\includegraphics[scale=0.65]{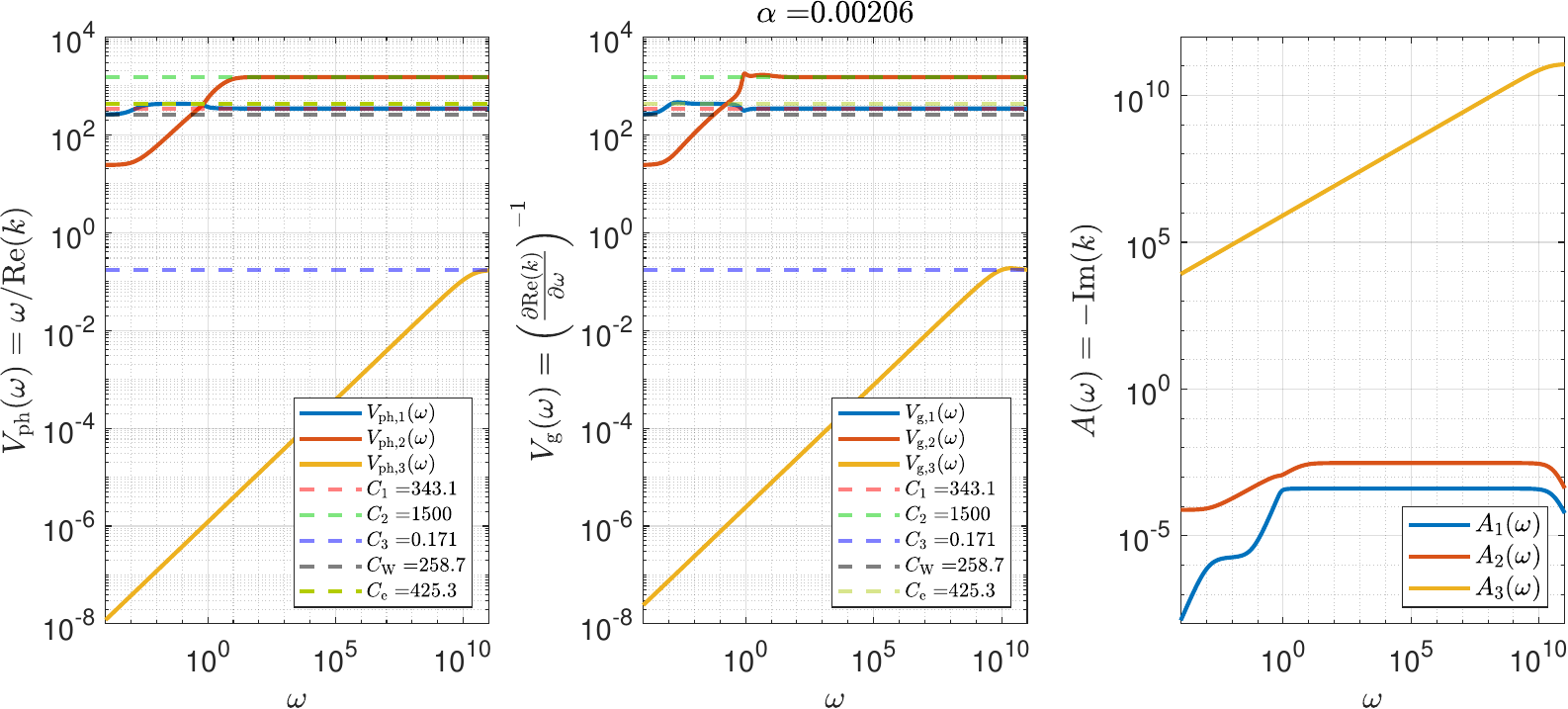}}
	\caption[Put some things in here]{Phase velocities (left), group velocities (middle), and 
			attenuation factor (right) versus the angular frequency $\omega = 2 \pi f$, where $f$ 
			is 
			the 
			frequency. The dashed lines shows the pure gas $C_1=\sqrt{\pd p_1(\rho_1,S)/\pd 
			\rho_1}$ and 
				liquid 
				$C_2=\sqrt{\pd p_2(\rho_2,S)/\pd\rho_2}$ characteristic 
				velocities, 
				equilibrium characteristic speed $C_{\text{e}}$, and Wood's sound speed 
				$C_{\text{W}}$. The mixture parameters are 
			$\alpha=0.00206$, $\beta = 1$, $\delta = 0.0025$, $\delta/\lambda = 10^{-8}$. The phase 
			pressures 
			parameters are extracted from the stationary bubble solution at the corresponding value 
			of 
			$\alpha$: $p_1 = 1.0132276 \cdot 10^{5}$, $p_2 = 1.0132500\cdot10^{5}$, while other 
			other 
			thermodynamic parameters were computed from the equations of state.}
	\label{fig.alpha0.002}
\end{figure}

The polynomial \eqref{eqn.poly} is a cubic polynomial on $k(\omega)^2$, and its roots correspond to 
three modes: the pressure modes of the two phases, and the capillarity mode associated with the 
surface 
tension. All three modes are stable as can be seen from the attenuation factors 
plotted in 
Fig.\,\ref{fig.alpha0.999}--\ref{fig.alpha0.002} and which are positive.

Figures \ref{fig.alpha0.999}--\ref{fig.alpha0.002} show typical dispersion curves for 
the thermodynamic parameters extracted from the stationary bubble solution corresponding to 
$\alpha=0.999$ (almost pure gas), $\alpha=0.519$ (mixed state), $ \alpha = 0.002$ (almost pure 
liquid).  The curves are plotted along 
side with the phase characteristic velocities $ C_a^2 = \frac{\pd p_a}{\pd \rho_a}$, the so-called 
equilibrium sound speed $C_\text{e}$ and Wood's sound speed $C_\text{W}$ (dashed lines) given by 
\cite{Gavrilyuk2002}
\begin{equation}
	C_{\text{e}}^2 = 
	\left( \rho \left( \frac{\alpha_1}{\rho_1 C_1^2} + \frac{\alpha_2}{\rho_2 C_2^2} \right) 
	\right)^{-1},
	\qquad
	C_{\text{W}}^2 = 
	 C_1^2 C_2^2 \left( \frac{\rho_1}{\alpha_1} + \frac{\rho_2}{\alpha_2} \right) 
	 \left(\frac{\rho_1 C_1^2}{\alpha_1} + \frac{\rho_2 C_2^2}{\alpha_2} \right)^{-1}.
\end{equation}

From these figures, one can note that Wood's speed $C_\text{W}$ serves as a low frequency limit for 
the sound speed of the gaseous phase $a=1$ (the light one), and the equilibrium speed $C_\text{e}$ 
is the high-frequency limit of the gaseous phase.

On the other hand, the liquid phase (the heavy one), has $C_2$ as the high-frequency limit, while 
its low-frequency limit is undetermined in general.

The most interesting behavior of the sound waves can be seen in Fig.\,\ref{fig.alpha0.002} 
corresponding to $\alpha=0.00206$ that can be considered as if a gaseous phase $\rho_1=1.205$ is 
dispersed in a heavy liquid phase $\rho_2=1000$. Thus, the velocity of the sound mode corresponding 
to the light phase may change from $C_\text{W}= 258.7$ at $\omega\to 0$ to $C_1=343.1$ at 
$\omega\to\infty$ through $C_\text{e}=425.3$ at moderate frequencies.

\section{Conclusion}

We have presented a new nonequilibrium \textit{two-pressure} diffuse interface model for two-fluid 
systems 
with surface tension. The model represents an extension of the SHTC two-fluid mixture model 
\cite{Rom1998,Rom2001,Romenski2007,RomDrikToro2010} in which the time and space gradients of the 
volume fraction 
are lifted to the level of state variables to account for spatial heterogeneities and microinertial 
effect in the vicinity of the diffuse interface. The resulting model belongs to the class of 
thermodynamically compatible hyperbolic equations for which both laws of thermodynamics hold.

The main difference from the existing diffuse interface approaches to surface tension such as 
\cite{Brackbill1992,Saurel2005,Schmidmayer2017,Chiocchetti2021} is that the pressure equilibrium 
condition $p_1=p_2$ is not assumed in our model. The pressure non-equilibrium is actually a 
fundamental assumption for introduction of the surface tension into the two-fluid SHTC model in 
order to gain the full SHTC structure of the governing equations. In particular, it is important 
for the variational and Hamiltonian formulations of the governing equations discussed in 
Sec.\,\ref{sec.app.VarPrinciple} and \ref{sec.generic}.

We presented two reduced relaxation limits of the model. In Sec.\,\ref{sec.relax.Limit.D}, the 
relaxation limit of the equations when the microinertia relaxation rate is very fast. In 
Sec.\ref{sec.relax.limit.all}, the relaxation limit for very fast pressure and velocity relaxation 
rates was derived that can be called single-velocity pressure-equilibrium model. These reduced 
models can be used for better understanding of the solution properties of the full system. In 
particular, we demonstrated that the Young-Laplace law is approximately holds in a mixture element 
as long as the gradient of the volume fraction is not equal to zero. Furthermore, the single-velocity 
pressure-equilibrium reduced model is a system of third-order partial differential equations which 
can be used to connect our model to some classical dispersive models. The dispersion relation of 
the full model can be obtained numerically and was studied in Sec.\ref{sec.dispers}. The full model 
has three modes: two pressure modes of the constituents and the third mode, related to the 
microinertia of the interface. The analysis revealed non trivial dependencies of the phase 
velocities on frequency. 

Although we could not conduct the hyperbolicity analysis of the full system completely, our 
preliminary study suggests that the full two-fluid model with surface tension is hyperbolic at 
least in the vicinity 
of equilibrium. Also, during this 
research, we found out that the energy potential is not strictly convex and thus formally the model does 
not have all properties of the SHTC class of equations. We shall address this issue in more detail in future.

Finally, we studied the stationary bubble solution, and demonstrated that the Young-Laplace law 
holds as a sharp interface limit for macroscopic interfaces. Although the new parameter $\delta$ is not a constant, from the theoretical 
viewpoint, and must be chosen as 
$\delta=\Vert \del \alpha \Vert^{-1}$, we demonstrated that in practical computations it can be 
taken as a precomputed constant, see \eqref{eqn.delta0}.

In future, we shall extend the model so that it applies also to complex fluids and solids, such as superfluid helium-4 or non-Newtonian fluids.

\section*{Acknowledgment}
The work of I.P. was financially supported via the Departments of Excellence Initiative 2018–2027 
attributed to DICAM of the University of Trento (grant L. 232/2016). Also, I.P. is a member of the 
INdAM GNCS group in Italy.
E.R. is supported by the Mathematical Center in Akademgorodok under the agreement No. 
075-15-2022-281 with the Ministry of Science and Higher Education of the Russian Federation.
M.P. was supported by Czech Science Foundation, project 23-05736S. M.P. is a member of the Nečas center for Mathematical Modelling.

\appendix

\section{Variational formulation}\label{sec.app.VarPrinciple}

Here we briefly recall the variational scheme that underlies every SHTC equations, see 
\cite{SHTC-GENERIC-CMAT}. The variational principle is formulated in the Lagrangian coordinates 
$\boldsymbol{X} = \left\{X^K\right\} $, $K=1,2,3$. For simplicity, we only consider the surface 
tension 
part.

Thus, we consider a general Lagrangian density $\Lambda = \Lambda(\alpha,\phi,\beta_K)$ which is a 
function of the volume fraction and their 
space and time gradients 
\begin{equation}\label{eqn.phi.beta}
	\phi:= \rmd_t \alpha, 
	\qquad
	\beta_K := \pd_K\alpha,
\end{equation}
where $\rmd_t = \rmd/\rmdt$ is the material time derivative and $\pd_K = \pd/\pd X^K$. Hence, 
varying the action integral
\begin{equation}
	\mathcal{A} = \int \Lambda(\alpha,\rmd_t \alpha,\pd_K\alpha) \rmdt 
	\rmd\boldsymbol{X}
\end{equation}
with respect to $\delta\alpha$, one immediately obtains the Euler-Lagrange equation
\begin{equation}
	\frac{\rmd \Lambda_\phi}{\rmdt} + \frac{\pd \Lambda_{\beta_K}}{\pd X^K} = \Lambda_{\alpha},
\end{equation} 
where $\Lambda_\alpha = \frac{\pd \Lambda}{\pd \alpha}$, $\Lambda_\phi = \frac{\pd \Lambda}{\pd 
\phi}$, and $\Lambda_{\beta_K} = \frac{\pd \Lambda}{\pd \beta_K}$. The Euler-Lagrange equation is 
formally a second-order PDE for $\alpha$. To obtain an extended first-order system for 
$\left\{\alpha,\phi,\beta_K\right\} $ as required by the SHTC theory, we need to 
provide an evolution equations for the vector $\beta_K$ and $\alpha$ which is in fact not a 
difficult 
task because these equations are trivial consequences of the definitions of $\beta_K$ and $\phi$. 
Indeed, the required system of first-order PDEs reads
\begin{subequations}\label{eqn.SHTC.Lagr1}
	\begin{align}
			\frac{\rmd \Lambda_\phi}{\rmdt} &+ \frac{\pd \Lambda_{\beta_K}}{\pd X^K} = 
			\Lambda_{\alpha},\\[2mm]
			\frac{\rmd \beta_K}{\rmdt} &- \frac{\pd \phi}{\pd X^K} = 0,\\[2mm]
			\frac{\rmd \alpha}{\rmdt} &= \phi.
	\end{align}
\end{subequations}

By introducing new potential as a partial Legendre transform of $\Lambda$
\begin{equation}
	U(\alpha,d,b_K):= \phi \Lambda_\phi - \Lambda,
\end{equation}
and new variables
\begin{equation}\label{eqn.d.def}
	d:=\Lambda_\phi, \qquad b_K:=-\beta_K,
\end{equation}
system \eqref{eqn.SHTC.Lagr1} can be rewritten as
\begin{subequations}\label{eqn.SHTC.Lagr2}
	\begin{align}
		\frac{\rmd d}{\rmdt} &+ \frac{\pd U_{b_K}}{\pd X^K} = 
		-U_{\alpha},\label{eqn.SHTC.Lagr2.d}\\[2mm]
		\frac{\rmd b_K}{\rmdt} &+ \frac{\pd U_d}{\pd X^K} = 0,\label{eqn.SHTC.Lagr2.b}\\[2mm]
		\frac{\rmd \alpha}{\rmdt} &= U_d,
	\end{align}
\end{subequations}
where we have used the standard properties of the Legendre transform
\begin{equation}
	U_\alpha = -\Lambda_\alpha,
	\qquad
	\phi = U_d.
\end{equation}

Finally, performing the Lagrange-to-Euler transformation of the energy potential, state variables, 
and time and space derivatives
\begin{subequations}
	\begin{align}
		&A^K_{\ i}:=\frac{\pd X^K}{\pd x^i},
		\qquad
		\frac{\pd}{\pd x^i} = A^K_{\ i}\frac{\pd }{\pd X^k},
		\qquad
		\frac{\rmd}{\rmdt} = \frac{\pd}{\pd t} + v^k \frac{\pd}{\pd x^k},\\[2mm]
		&E:= \det(\boldsymbol{A}) U,
		\qquad
		d:= \det(\boldsymbol{A})d, 
		\qquad
		b_i := A^K_{\ i}b_k,\\[2mm]
        &\frac{\rmd}{\rmdt}A^K_{\ i}+A^K_{\ j} \frac{\pd v^j}{\pd x^i} =0
 \end{align}	
\end{subequations}
as detailed in 
\cite{SHTC-GENERIC-CMAT}, equations 
\eqref{eqn.SHTC.Lagr2} become \eqref{eqn.PDE.adb}. 
Note that, in \eqref{eqn.SHTC.Lagr2.d}, an equivalent of the source term $\lambda\rho E_d$ 
in \eqref{d_eqn} is missing because it is a dissipative term and cannot be included 
into the variational scheme but must be added afterwards in correspondence with the second law of 
thermodynamics \cite{SHTC-GENERIC-CMAT}.

\section{Hamiltonian formulation}\label{sec.generic}
Equations \eqref{MasterSysSF} consist of a reversible part and an irreversible part 
(the two source terms with the rates $\chi^{-1}$ and $\ceps^{-1}$). The purpose of this Section is 
to show that the 
reversible part has a Hamiltonian structure, meaning that it is generated by a Poisson bracket. 
This underlines the internal consistency of the equations and provides to them an alternative 
geometric interpretation \cite{PKG_Book2018}. 

Poisson brackets are a cornerstone of analytical mechanics \cite{Goldstein}, where they typically 
come out of variation of action. Then, 
they come in the so-called canonical form for the pairs of coordinates and their respective 
momenta. But in continuum 
mechanics, which is the case here, Poisson brackets are typically non-canonical, they are 
degenerate, and do not need to have state variables in the form of pairs of coordinates and momenta 
\cite{clebsch,arnold,dv,PKG_Book2018}. 

In general, Poisson brackets are bilinear operators from a space of functionals of some state variables $\yy$ to 
the same space. Moreover, they are skew-symmetric, $\{F,G\} = -\{G,F\}$, for each two 
functionals $F(\yy)$ and $G(\yy)$, and they satisfy the Leibniz rule and Jacobi identity. The state variables can be 
for 
instance positions and momenta, which is the usual case in analytical mechanics, but also fields as 
density, momentum density, and others. The Leibniz rule, 
$\{F,GH\} = \{F,G\}H + G\{F,H\}$ means that Poisson brackets behave as derivatives, which in particular means that the resulting evolution equations do not depend on constant shifts of energy. Finally,  
Jacobi identity, $\{F,\{G,H\} + \{G,\{H,F\} + \{H,\{F,G\} = 0$, expresses self-consistency of 
Hamiltonian mechanics \cite{fecko,PPK2020}. Once the Poisson bracket for a state variable vector $\yy$ 
is 
determined, Hamiltonian evolution of $\yy$ is given by 
\begin{equation}\label{eq.ham}
\dot{\yy} = \{\yy,\mathcal{E}\},
\end{equation}
where $\mathcal{E} = \int e(\yy) d\xx$ is the total energy of the system and $e(\yy)$ the volumetric energy density.

Now, we approach the Hamiltonian structure of Equations \eqref{MasterSysSF}, which can be seen as 
evolution equations for state variables $\rho$ (mass density), $\mm$ (momentum density), 
$\crho=\rho c$ (density of a mixture component), $\ww$ (relative velocity), $\alpha$ (volume 
fraction), $\DD = \rho D$ (volume fraction rate density), $\BB$ (interface gradient), and $s = \rho 
S$ (entropy density per volume).
Fields $\rho$, $\mm$, and $s$, which describe fluid mechanics, are known to have Hamiltonian evolution generated by Poisson bracket
\begin{multline}
\{F,G\}^{(\mathrm{FM})} = \int \rho \left(\partial_i F_{\rho} G_{m_i}-\partial_i G_{\rho} 
    F_{m_i}\right)d\xx 
    +\int m_i \left(\partial_j F_{m_i} G_{m_j}-\partial_j G_{m_i} 
    F_{m_j}\right)d\xx 
    +\int s \left(\partial_i F_{s} G_{m_i}-\partial_i G_{s} 
    F_{m_i}\right)d\xx,
\end{multline}
where FM stands for fluid mechanics \cite{arnold,PKG_Book2018}. Note that in this section the 
subscripts stand for functional derivatives, not partial derivatives as in the preceding sections. 
From the geometric point of view, this bracket expresses that the velocity field $\vv = 
\frac{\partial e}{\partial \mm}$ advects itself as well as two scalar densities, $\rho$ and $s$. 
From the algebraic point of view, the bracket expresses dynamics on the Lie algebra dual of a 
semidirect product \cite{marsden-weinstein, esen2012geometry}.  

Another part of Equations \eqref{MasterSysSF} consists of the equations for fields $\crho$ and 
$\ww$ (density of a species and its relative velocity with respect to the other species). These 
fields form a cotangent bundle and the whole cotangent bundle is then advected by the overall 
velocity $\vv$, which gives the mixture part of the overall bracket
\begin{align}
    \{F,G\}^{(\mathrm{mixture})}=& \{F,G\}^{(FM)}\nonumber\\
        &+\int \cbar \left(\partial_i F_{\cbar} G_{m_i}-\partial_i G_{\cbar} 
    F_{m_i}\right)d\xx \\
    &+\int \left(\partial_i F_{\cbar} G_{w_i}-\partial_i G_{\cbar} 
    F_{w_i}\right)d\xx\nonumber\\
    &-\int \partial_i w_j \left( F_{w_j} G_{m_i}-G_{w_j} 
    F_{m_i}\right)d\xx\nonumber\\
    &+\int w_i \left(\partial_j F_{m_i} G_{w_j}-\partial_j G_{m_i} F_{w_j}\right)d\xx,
\end{align}
see \cite{SHTC-GENERIC-CMAT} for more details.

Then, the pair of fields $\{\BB,\DD\}$ have analogical dynamics as the pair $\{\ww,\crho\}$, which 
leads to an analogical 
part of the Poisson bracket. However, density $\DD$ is also part of another cotangent bundle, this 
time with function $\alpha$, resulting in their canonical coupling,
\begin{equation}
\{F,G\}^{(\delta)} = \frac{1}{\delta}\int (F_{\DD} G_{\alpha}- G_{\DD} F_{\alpha}) d\xx,
\end{equation}
where $\delta(\alpha)$ is an adjustable functional of $\alpha$. Note that in order to keep validity 
of the Jacobi identity, $\delta$ can not depend on other state variables than $\alpha$. Since 
$\alpha$ is a function, not a density (as opposed to for instance $\rho$), it is advected by terms
\begin{equation}
\int \left(\pd_i F_{\alpha}G_{m_i}-\pd_i G_{\alpha}F_{m_i}\right) d\xx
\end{equation}
in the resulting Poisson bracket.

The overall Poisson bracket for state variables in Equations \eqref{MasterSysSF} is then 
\begin{align}\label{eq.bracket}
\{F,G\}^{(\mathrm{surface\,tension})} =&\{F,G\}^{(\mathrm{mixture})} + \{F,G\}^{(\delta)}+\int \left(\partial_i F_{\alpha}G_{m_i}-\partial_i G_{\alpha}F_{m_i}\right) d\xx\nonumber\\
&+\int \DD \left(\partial_i F_{\DD} G_{m_i}-\partial_i G_{\DD} F_{m_i}\right)d\xx \nonumber\\
&+\int \left(\partial_i F_{\DD} G_{B_i}-\partial_i G_{\DD} 
    F_{B_i}\right)d\xx\nonumber\\
    &-\int \partial_i B_j \left( F_{B_j} G_{m_i}-G_{B_j} 
    F_{m_i}\right)d\xx\nonumber\\
    &+\int B_i \left(\partial_j F_{m_i} G_{B_j}-\partial_j G_{m_i} F_{B_j}\right)d\xx,
\end{align}
where validity of Jacobi identity was checked by program \cite{kroeger2010}. From the geometric point of view, this Poisson bracket expresses evolution three-forms $\rho(t, \xx)\oomega$ and $s(t,\xx)\oomega$, where $\oomega = dx\wedge dy\wedge dz$ is the volume three-form \cite{fecko}, advected by the overall velocity that also advects the momentum density $m_i(t, \xx)dx^i \otimes \oomega$. Moreover, three-form $\rho(t,\xx)c(t,\xx)\oomega$ is coupled with one-form $w_i(t,\xx)dx^i$ (differential of the dual function to that three-form), and both are advected by the velocity.
Finally, the three-form $\rho(t,\xx)D(t,\xx) \oomega$ with its canonically coupled dual element, zero-form or function $\alpha(t,\xx)$, are advected by the velocity. Three-form $\rho(t,\xx)D(t,\xx) \oomega$ is, moreover, coupled with one-form $B_i(t,\xx)dx^i$, which is constructed as the differential of the dual to that three-form, and both are advected (Lie-dragged) by the velocity field. 

Once energy of the system is determined, Hamiltonian evolution \eqref{eq.ham} with Poisson bracket 
\eqref{eq.bracket} leads to the reversible (non-dissipative) part of Equations \eqref{MasterSysSF}.

\printbibliography

\end{document}